\documentclass[sigconf]{acmart}

\AtBeginDocument{%
  }

\copyrightyear{2026}
\acmYear{2026}
\setcopyright{cc}
\setcctype{by}
\acmConference[CHI '26]{Proceedings of the 2026 CHI Conference on Human Factors in Computing Systems}{April 13--17, 2026}{Barcelona, Spain}
\acmBooktitle{Proceedings of the 2026 CHI Conference on Human Factors in Computing Systems (CHI '26), April 13--17, 2026, Barcelona, Spain}
\acmPrice{}
\acmDOI{10.1145/3772318.3791792}
\acmISBN{979-8-4007-2278-3/2026/04}

\begin{document}

\title{How do the Global South Diasporas Mobilize for Transnational Political Change?}

\author{Dipto Das}
\affiliation{
    \department{Department of Computer Science}
    \institution{University of Toronto}
    \city{Toronto}
    \state{Ontario}
    \country{Canada}
}
\email{dipto.das@utoronto.ca}

\author{Afrin Prio}
\affiliation{
    \department{Department of Engineering Science}
    \institution{University of Toronto}
    \city{Toronto}
    \state{Ontario}
    \country{Canada}
}
\email{afrin.prio@mail.utoronto.ca}

\author{Pritu Saha}
\affiliation{
    \department{Department of Clinical Psychology}
    \institution{University of Dhaka}
    \city{Dhaka}
    \country{Bangladesh}
}
\email{pritusaha22@gmail.com}

\author{Shion Guha}
\affiliation{
    \department{Faculty of Information}
    \institution{University of Toronto}
    \city{Toronto}
    \state{Ontario}
    \country{Canada}
}
\email{shion.guha@utoronto.ca}

\author{Syed Ishtiaque Ahmed}
\affiliation{
    \department{Department of Computer Science}
    \institution{University of Toronto}
    \city{Toronto}
    \state{Ontario}
    \country{Canada}
}
\email{ishtiaque@cs.toronto.edu}


\begin{abstract}
  This paper examines how non-resident Bangladeshis mobilized during the 2024 quota-reform turned pro-democracy movement, leveraging social platforms and remittance flows to challenge state authority. Drawing on semi-structured interviews, we identify four phases of their collective action: technology-mediated shifts to active engagement, rapid transnational network building, strategic execution of remittance boycott, reframing economic dependence as political leverage, and adaptive responses to government surveillance and information blackouts. We extend postcolonial computing by introducing the idea of \emph{``diasporic superposition,"} which shows how diasporas can exercise political and economic influence from hybrid positionalities that both contest and complicate power asymmetries. We reframe diaspora engagement by highlighting how migrants participate in and reshape homeland politics, beyond narratives of integration in host countries. We advance the scholarship on financial technologies by foregrounding their relationship with moral economies of care, state surveillance, regulatory constraints, and uneven international economic power dynamics. Together, these contributions theorize how transnational activism and digital technologies intersect to mobilize political change in Global South contexts.
\end{abstract}



\maketitle

\noindent\textbf{Content Warning:} This paper discusses state violence, surveillance, and protest-related deaths.

\section{Introduction}\label{sec:introduction}
More than 304 million people live outside their countries of birth~\cite{paez2025top}. About 72\% of these migrants originate from the Global South, moving either to Global North countries or within other Global South regions~\cite{horwood2021repositioning}. Last year, these diasporic communities sent over \$650 billion as remittances, i.e., funds from the country of work back to a home country, which are often low- and middle-income nations in the Global South~\cite{ratha2024remittance, worldbank2024remittances} to sustain families, economies, and cultures while maintaining enduring bonds to their homelands. These remittances often circulate through both formal channels (e.g., banks, mobile money) and extensive informal networks (e.g., hundi) as migrants navigate various financial and regulatory constraints~\cite{rahman2014social}. However, human-computer interaction (HCI) and social computing research have primarily examined migrants and diasporas in terms of integration in their host countries: how migrants use technologies to access services, build communities, or adapt to unfamiliar social, cultural, and political systems in their new countries of residence~\cite{sabie2022decade, hsiao2018technology, dosono2018identity}. While digital technologies often emerge as primary channels through which the diasporic communities remain informed about and participate in the discourse around the contemporary sociopolitical events of their home country~\cite{brinkerhoff2009digital, bhattacharjee2025residual}, there is a dearth of CHI scholarship that looks into how diasporas mobilize as direct political actors in their homeland politics through those platforms. We take up this gap by asking: How do the Global South diasporas use digital platforms and remittances to advance transnational political change?

HCI and social computing literature has examined how digital platforms facilitate collective action, support resilience, and play vital roles in protests, activism, and political discourse worldwide~\cite{bilic2025digital, milan2015algorithms, westerhuis2022politics, postmes2002collective}. These platforms become even more critical in contexts where political institutions are repressive, corrupt, mistrusted, and resistant to democratic participation by enabling grassroots actors to coordinate across scales, amplify visibility, and challenge power~\cite{wilson2021cross, rahimi2011agonistic, reuter2015online}. However, most research emphasizes on-the-ground participants located at the site of struggle and only a few pay little attention to the diasporic actors~\cite{ansar2024digital, moss2016transnational, moss2022arab}. Similarly, as HCI research on financial technologies (FinTech) predominantly focuses on innovative technologies and personal finances across various demographics~\cite{bitrian2021making, dai2025envisioning, chiang2018exploring}, it pays comparatively less attention to remittances--particularly their role as instruments of political resistance--leaving the financial leverage of diasporas in homeland politics insufficiently understood.

Whereas prior work has separately explored online activism~\cite{erete2021can, das2022understanding}, diasporic identity work~\cite{brinkerhoff2009digital, dosono2020decolonizing}, and remittance transfers through FinTech~\cite{chang2023migration, rohanifar2021money}, we argue that their entanglement can generate new forms of political leverage that warrant investigation. The remittance boycott--defined as the collective and coordinated withholding or rerouting of diaspora earnings away from state-aligned financial institutions and technologies--by the non-resident Bangladeshis (NRBs)~\cite{mavis2024remittance} during the quota reform movement in July 2024 that later turned into a pro-democracy uprising and eventually toppled the authoritarian regime of the last 16 years~\cite{unohchr2024preliminary} makes a strong case to illustrate this. During that movement, the NRBs transformed remittance from a familial economic practice into an instrument of collective political resistance, enabled by online platforms that mediated visibility, coordination, and trust. In this context, digital platforms played multiple roles: social media enabled information flow and coordination, while financial technologies, especially remittance channels, mediated economic leverage and diasporic influence. Based on a qualitative study informed by semi-structured interviews with NRBs residing in North America, Europe, and the Middle East, we examine their diasporic agency by tracing how they mobilized social media and remittance infrastructures as interconnected channels of transnational political action.

It is important to note that while our work adhered to the interpretivist tradition, the context of the work and its focus on the July 2024 movement in Bangladesh are not only inherently political but also temporal. We have identified four phases of the NRB's mobilization in boycotting remittance during this movement: (1) becoming politicized as acts of resistance against political intimidation and through emotionally charged visuals circulating online; (2) building weak-tie networks across transnational boundaries; (3) executing a remittance boycott by improvising informal financial channels; and (4) navigating barriers such as surveillance and a nationwide internet blackout. Our findings contribute to CHI and social computing scholarship in three ways. First, we extend critical computing by theorizing \emph{``diasporic superposition"} as a way of understanding hybrid positionalities in postcolonial computing, illustrating how diaspora and immigrants navigate overlapping positions of privilege and vulnerability. Second, we expand migration studies by foregrounding how diasporas are not only recipients of host-country policies but also active transnational agents who operate through moral economies and are capable of shaping homeland politics. Third, we inform FinTech research and design by situating remittance within care and resistance, reflecting on how their designs can be appropriated for both survival and protest. Together, these contributions theorize how digital platforms, diasporic communities, and economic instruments intersect to organize transnational collective action for political change in the Global South contexts.
\section{Literature Review}
This section reviews three strands of scholarship that frame our study. First, we describe prior research on digital platforms and collective action and how that emphasizes on-the-ground activism while sidelining diasporic roles. Second, we turn to interdisciplinary research on digital diasporas and their transnational agency. Finally, we review scholarship on remittance-related financial technologies (FinTech), highlighting a gap in understanding how remittances embody power dynamics, care, and instruments of contestation.

\subsection{Digital Platforms and Collective Action in Crisis and Protest}
For more than a decade, HCI and computer-supported cooperative work (CSCW) researchers have studied how digital platforms enable and constrain distributed publics to pursue collective action toward shared goals~\cite{olson1971logic, li2018out, bennett2012logic, enjolras2013social, schradie2018digital}. A key insight from this research is that crisis and unstable contexts, such as political unrest and protests, make their role especially visible: individuals who could not safely or feasibly engage offline can still express solidarity, contribute funds, share situational updates, or maintain live reporting~\cite{starbird2011voluntweeters, vieweg2010microblogging, li2019people, al2010blogging}. During the 2021 Henan floods, for instance, online users organized information and support in ways that underscored the indispensability of online collective action and the need for mechanisms that manage overload at scale~\cite{chen2025navigating}. These platforms can also attenuate the hesitation to act in certain parts of the community by making aggregate participation legible. Besides surfacing endorsement, debate, and commitments~\cite{shaw2014computer}, crowdfunding sites' real-time progress bars and backer counts, converting private intention into public momentum~\cite{cheng2014catalyst, hallam2016internet}.

Most platforms offer various interactional mechanisms that collectively scaffold action: posting, commenting, liking, sharing, livestreaming, group formation (e.g., Facebook Groups), and conversation organization (e.g., hashtags)~\cite{li2025explaining, barron2022quantifying}. Such interactions knit together weak-tie networks~\cite{granovetter1973strength, aral2016future}, connecting individuals who do not share strong, frequent contact in their everyday lives. Empirical work has measured how such weak ties in Facebook Groups sustain information exchange and efficacy through routine micro-engagement~\cite{baborska2016exploring}. In the context of Bengali communities, these interactional practices extend the cultural form of adda--informal political discussion--to digital spaces~\cite{das2025btpd}, where it both translates and reconfigures familiar modes of collective deliberation~\cite{das2024reimagining, chakrabarty2009provincializing}. Although the ties formed on messaging platforms (e.g., Telegram) are harder to observe directly, group-level metrics such as replies, forwards, mentions, and cross-posts prove emergent connectivity~\cite{wischerath2024spreading}. Rapid weak-tie formation reduces degrees of separation, accelerates diffusion, and allows organizers to reach into everyday life at speed. Algorithmic ranking, moderation, and adversarial reporting can suppress or distort activist messages; state power can intervene more bluntly through platform throttling or information blackouts~\cite{reyes2023those}. Consequently, activists adopt strategies to exploit platform logics--strategic hashtagging, viral imagery, ephemeral stories, and migration across channels--to preserve visibility and safety~\cite{barron2022quantifying,shaw2014computer}.

In crisis informatics, users document unfolding events, coordinate volunteers, and distribute resources when formal institutions are slow or mistrusted~\cite{shaw2014computer, greijdanus2020psychology}. The net result is a socio-technical pattern: low-barrier contribution and algorithmic amplification, braided through weak ties and group structures, enable large, distributed publics to act collectively on compressed timescales~\cite{cheng2014catalyst, vieweg2010microblogging}. While technologies are often critiqued for further marginalizing underrepresented communities in terms of race~\cite{das2023decolonization}, gender~\cite{nova2019online, nova2021facebook, sambasivan2019they}, ethnicity~\cite{rifat2024politics, sultana2024civics}, linguistic norms~\cite{das2021jol}, religion~\cite{das2021jol, rifat2024politics}, economic class~\cite{sambasivan2018privacy}, rurality~\cite{sultana2022toleration}, and literacy~\cite{hasan2021oshudh} in the Global South contexts like Bangladesh, the discrimination across different intersectional Bengali identities exacerbates with the increasing adoption of contemporary algorithmic systems like large language models and datasets~\cite{das2025auditing, das2024colonial, das2025datasets, das2023toward}. However, digital platforms also sustain participation in political discourse around long-term, normalized crises that are less visible and not confined to a single geographical area~\cite{das2022collaborative, das2022understanding}. These crises affect communities as dispersed collectives rather than bounded locations. For example, Das and colleagues show how Bengali ethnolinguistic communities across multiple countries engage in collaborative reclamation of narrative agency and reimagine sociopolitical and economic practices to resist the enduring crisis of colonialism~\cite{das2022collaborative, das2024reimagining}. Similarly, the diaspora mobilizations surrounding the 2020–2021 Indian farmers' protests demonstrated how online platforms united communities connected through moral obligation to agrarian professions and shared political commitments across Canada, the US, and India, synchronizing frames, logistics, and fundraising~\cite{monteiro2021farmer}. This latter body of literature highlights two key assumptions that continue to shape dominant narratives in HCI research on digital protests that activism is geographically bound to the ``site" of activism, and that the vital actors are those navigating immediate, on-the-ground risks. Such perspectives relegate diasporic actors to auxiliary roles--fundraisers, donors, amplifiers, or sympathetic audiences--rather than recognizing them as constitutive agents of political struggle. Hence, crisis informatics literature should extend beyond localized protest contexts to examine the transnational technologies and practices through which diasporas mobilize. The next subsection elaborates on this turn by foregrounding diasporic technologies and transnational agency.

\subsection{Digital Diasporas' Resource Leverage and Transnational Political Agency}
Online spaces are crucial sites for diasporic communities to maintain existing cultural continuity and forge new social bonds~\cite{brinkerhoff2009digital, alonso2010diasporas}. Thus, digital platforms function not just as communication channels but emerge as a vital medium to foster the sense of belonging and play broader sociopolitical role, giving rise to the \emph{digital diaspora}--a transnational community that uses digital technologies to sustain identity, solidarity, and political engagement across borders~\cite{ponzanesi2020digital, candidatu2019digital}.

HCI researchers have examined the experiences of migrants, refugees, and immigrant communities, highlighting how technologies mediate their adaptation, inclusion, and well-being in new sociopolitical environments, particularly focusing on their integration into host societies, whether through improving access to services, enabling communication with local communities, or addressing barriers of language, literacy, and digital exclusion~\cite{sabie2022decade, batool2024gendered}. For example, studies have explored how immigrants navigate immigration processes~\cite{chen2025navigating}, utilize information and communication technologies (ICTs) to build social capital in resettlement contexts~\cite{hsiao2018technology}, maintain their mental health~\cite{ayobi2022digital}, and deliberate on political topics of their country of residence~\cite{dosono2018identity}. Moving beyond these the lenses of vulnerability, adjustment, and assimilation, contemporary scholarship have theorized diasporas as transnational political actors and studied how they engage in protest, advocacy, and contestation through information diffusion, motivational reinforcement, and networked coordination across borders~\cite{jost2018social}. In the case of the Arab Spring, Libyan, Syrian, and Yemeni diasporas mobilized while physically being in the US and the UK and emerged as crucial actors in sustaining political contention when domestic movements faced severe repression~\cite{moss2016transnational, moss2022arab}. Recent studies described how the Rohingya diaspora used Facebook to contest ethnic marginalization in Myanmar~\cite{ansar2024digital}, emphasizing digital diasporas' role in homeland politics.

A parallel body of work underscores how diasporic resources--financial, social, and political--enable transnational agency. Migrants circulate not only money but also ideas, practices, and norms that reshape political life in their homelands~\cite{lacroix2016social, krawatzek2020two}. For example, Saxenian illustrates how highly skilled migrants circulated knowledge and capital between Silicon Valley and their home countries, reshaping regional economies~\cite{saxenian2006new}. States have also mobilized diasporas through financial instruments such as diaspora bonds, as in the cases of Israel and India, where governments tapped expatriate communities to secure funds during periods of economic or political crisis~\cite{chander2001diaspora, gevorkyan2021can}. However, political economy research focuses more on the influence of diasporas on political sentiment at large and has shown that household transfers from abroad can reduce dependence on state assistance, thereby lowering the risks of dissent but increasing the likelihood of protest in authoritarian contexts~\cite{escriba2018remittances}. This duality can be explained by the scale and distribution of these transfers, which shape support for different forms of protest~\cite{lopez2025migrant} and exhibit a non-linear relationship in which transfers dampen contention at low levels but facilitate it once they surpass a critical threshold~\cite{aher2025diaspora}. Beyond households, diasporas can crowdfund and channel resources into contentious movements and to sustain resistance against non-democratic regimes~\cite{flanigan2017crowdfunding, international2022crowdfunding}.

Contemporary events highlight how diasporas use not only remittance inflows but also remittance boycotts as instruments of political leverage. Ethiopian diaspora groups threatened to withhold remittances as a protest against state violence and human rights abuses~\cite{borkena2018ethiopian}. A similar routing of financial leverage in shaping conflict and peacekeeping was also demonstrated by the Somali diaspora politics~\cite{horst2008transnational}. In the context of our study, non-resident Bangladeshis (NRBs) boycotted formal remittance channels and rerouted funds through informal networks to signal their disapproval of government repression.

Such tactics invert the conventional narrative of remittances as apolitical family support, showing how diasporas consciously reconfigure financial leverage while being embedded in a moral economy of obligation, solidarity, and resistance. Moral economy refers to collective beliefs, values, and norms that shape economic interactions, emphasizing mutual obligations and fairness over purely market-driven interest~\cite{thompson1971moral, srinivasan2020myths}. From this perspective, decisions to contribute--or withhold--resources are framed as ethical stances toward state injustice, highlighting how remittances operate simultaneously as systems of transfer, care, and contestation. As the accounts of remittance boycotts in CHI and social computing scholarship remain sparse, our paper aims to address that gap.


\subsection{Remittance as Sociotechnical Arrangements of Care and Contestation}
As FinTech increasingly mediates people's everyday finances, HCI and CSCW research has explored various technologies (e.g., blockchain, budgeting application) and different aspects of users' practices (e.g., motivation, experience, trust, risk perception)~\cite{frohlich2022blockchain, bitrian2021making}. A large portion of this literature focuses on FinTech designs and personal finance strategies~\cite{chiang2017understanding, sowon2024role, bitrian2021making} across various age, gender, and geographic demographics~\cite{kaewkitipong2022human, dai2025envisioning, smith2025gendered, centellegher2018mobile}. Beyond that, FinTech research in the Global South has also focused on the financial systems through which migrants and their families send and receive remittances.

Mobile money platforms such as M-Pesa in Kenya have become emblematic of the transformative potential of digital finance, enabling rapid, secure, and relatively low-cost transfers in regions with limited access to traditional banking institutions~\cite{maurer2012mobile}. These systems are often studied as success stories where remittances have bypassed the need for extensive banking networks and broadened financial inclusion~\cite{donovan2012mobile}. Since blockchain-based systems could reduce fees, increase transparency, and bypass state-controlled monetary reserves~\cite{narayanan2016bitcoin}, researchers have explored it as another disruptive alternative to centralized remittance channels~\cite{chiang2018exploring} and have found that migrants' willingness to adopt blockchain-based systems depends less on the technical novelty of distributed ledgers and more on whether platforms make money flows legible to senders and recipients. Hence, such solutions remain experimental and have not displaced entrenched incumbents, such as online money transfer platforms (e.g., Western Union, MoneyGram) or large banks, despite their high fees (average 4.1\% for the former compared to 12\% for the latter)~\cite{sohst2024leaving}.

HCI has also paid attention to the persistence of informal remittance systems, such as \emph{hundi} and \emph{hawala}~\cite{rohanifar2021money, maimbo2003informal}. Hawala is a broad Middle Eastern system of cross-border trust-based money transfers, where hawaladars (brokers) balance transactions through informal credit and trade networks rather than through formal banking rails~\cite{martin2009hundi}. Hundi is its South Asian variant, historically utilizing written notes. While often criminalized by states and often regarded as an opprobrious marker of the black market economy~\cite{sharma2025coinfused, martin2009hundi}, migrants rely on these channels for speed, lower costs, and social embeddedness. Such informal methods and ``workarounds" persist because formal platforms fail to meet users' contextual needs~\cite{sowon2024role}.

Despite the extensive technocratic literature, computing and design scholarship rarely conceives of remittances as political instruments. Other disciplines, such as development studies, approach remittances from that angle and often position it as a vital lifeline for Global South economies~\cite{ratha2024remittance}. For many countries--including Bangladesh, the Philippines, and Nepal--remittances account for more than 5-30\% of national GDP~\cite{worldbankgroup2024personal}, stabilizing macroeconomic conditions while providing direct household income. At the household level, remittances often provide for basic needs: food, housing, healthcare, and education~\cite{ifad2025reasons}. Development economists have celebrated remittances as countercyclical flows that sustain communities during crises such as natural disasters or political unrest~\cite{ansar2024digital, escriba2018remittances}. Critical scholarship complicates this optimistic narrative by identifying the disproportion where women are either senders or recipients of remittances~\cite{parrenas2005children} and how the remittance ecosystem privileges migrants with access to literacy, smartphones, and reliable networks~\cite{sabie2022decade}. These critiques conceptualize remittance flows not only as financial transactions but also as social and moral practices situated within family and community life.

The asymmetries in remittance systems that sustain the economies of the Global South, but are often mediated by institutions and regulations of the Global North, foreground postcolonial concerns about power-laden design and uneven economic power, as outlined by~\cite{irani2010postcolonial}. As some remittance receiver countries become reliant on external inflows rather than building robust domestic industries, this dynamic reproduces economic dependency. Moreover, the centralization of remittance ecosystems within state-linked foreign exchange reserves enables authoritarian regimes to exploit migrants' contributions for political ends. Therefore, we argue that remittances and associated FinTech should be viewed as contested sociotechnical arrangements embedded in uneven global hierarchies of capital and governance. The remittance boycott by NRBs demonstrates how diasporic communities can transform remittances from a purely economic transaction into an instrument of dissent. Our study seeks to inform FinTech design research of these political dimensions.

\section{Research Context}
\begin{figure}[!ht]
    \centering
    \includegraphics[width=\linewidth]{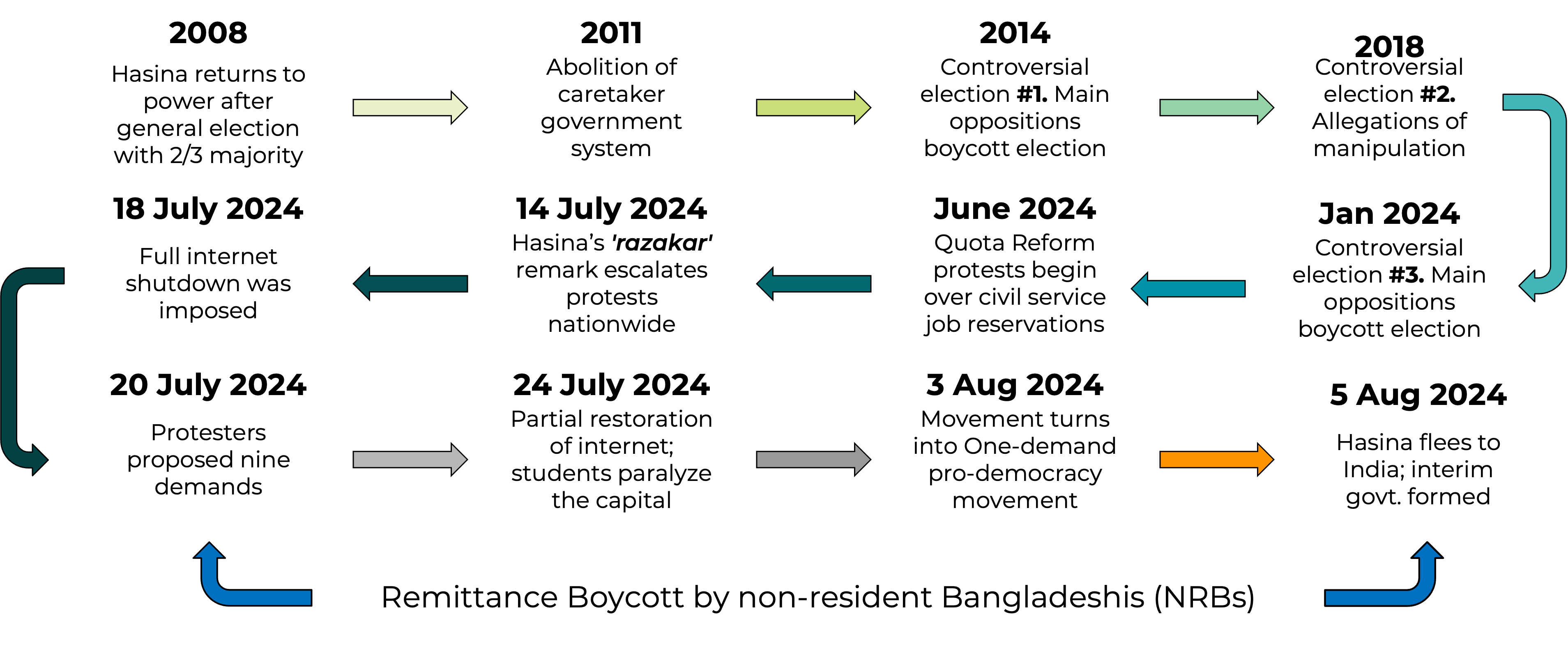}
    \caption{Timeline of Bangladesh's recent political past leading to the 2024 uprising and tentative timeline of the remittance boycott.}
    \label{fig:timeline}
\end{figure}

To broaden the readership of this paper, this section contextualizes the 2024 quota-reform turned pro-democracy movement in Bangladesh and the associated remittance boycott within the country's recent political history. Figure~\ref{fig:timeline} shows a timeline of recent and past events directly related to these developments. After providing a brief overview of the country's long history of political protests, we explain this timeline in sequence, detailing the consolidation of power under the Awami League (AL) party leader, Sheikh Hasina, and the series of events that shaped the 2024 movement — dynamics that repeatedly appeared in our study.

\subsection{Bangladesh's Political Protest Culture and the Road to 2024}
Bangladesh's contemporary protest culture is profoundly shaped by a history of authoritarianism and resistance against it. The country's last explicit dictatorship ended in 1990, when a military-ruled government was ousted by a rare alliance between AL and its historical rival, the Bangladesh National Party (BNP)~\cite{baxter1991bangladesh}. Throughout the following years, Bangladeshi politics remained polarized, with the AL and BNP taking turns at the helm, until Hasina returned to power after the 2008 general election with a more than two-thirds majority in parliament~\cite{panday2010analyzing}. While the first few years of Hasina's rule witnessed technological and economic development~\cite{al2013applications}, in 2011 her AL-led government exploited its parliamentary majority to abrogate the constitutional provision for non-partisan interim governance for elections~\cite{prodip2014abolition}. With heightened political distrust, polarization, and corruption in Bangladesh, this led to significant suppression of opposition voices and allegations of manipulation in the 2014, 2018, and 2024 elections~\cite{moniruzzaman2019regularization, riaz2021anatomy, mollah2018parliamentary}. Over time, these tendencies resulted in an increasingly autocratic political order with a shrinking space for opposition, government control over speech and assembly, and exploitation of law enforcement agencies to target critics~\cite{md2025awami, surie2023glass}. Throughout the AL's long tenure since 2008, the country has experienced multiple waves of political mobilization surrounding various issues (e.g., quota reform, road safety)~\cite{prodip2014abolition, tanjeem20232018}.

Political mobilization in Bangladesh has typically been characterized by a repertoire of disruptive tactics ranging from strikes (\textit{hartals}) and blockades to marches, sit-ins, hunger strikes, and, in more violent moments, street clashes and bombings~\cite{hossain2017significance, jackman2021students}. Universities have traditionally been the hotbeds of such activity. While there are independent student politics, most major political parties maintain affiliated student wings that they use to shape and control campus politics~\cite{jackman2021students}. With AL in power, the branches of its student wing, \textit{Chatro League} (CL), have dominated campus politics for decades, generally utilizing campuses as de facto partisan fiefdoms and exercising political influence far beyond university campuses. Over time, the government's systematic crackdown on legitimate opposition not only suppressed mass mobilization, even amid rising public discontent, but also steadily drained the broader political protest ecosystem~\cite{jackman2019threat, jackman2021students, ghosh2023conceptualizing}. These constraints gradually increased tensions, particularly among the students, who remained one of the few groups to possess both organizational power and historical legitimacy as drivers of political change, questioning the authoritarian regime and undermining the ruling party's legitimacy.

\subsection{The 2024 Protests: From Quota Reform to Democracy}
One such example was the Quota reform movement of 2024. It began in early June as a relatively contained, non-violent protest over civil service job quotas, since more than 56\% of public jobs in Bangladesh were reserved for individuals with different preferential allocations~\cite{das2024bangladesh}, 30\% of which went to the children and grandchildren of freedom fighters in the Bangladesh Liberation War of 1971~\cite{unohchr2024preliminary}. As the students demanded to bring such preferential quotas down to a lower number, Hasina rhetorically questioned in a press conference on July 14 whether the descendants of \textit{``razakars''} (a derogatory term for those who colluded against the country's independence and are widely regarded as traitors) should be favored instead~\cite{banka2024amidst}. As people were provoked by her statement, which dichotomized them as either descendants of the war veterans or the Razakars, the quota reform movement quickly escalated into a nationwide, anti-government uprising.

\begin{figure}[!ht]
    \centering
    \includegraphics[width=\linewidth]{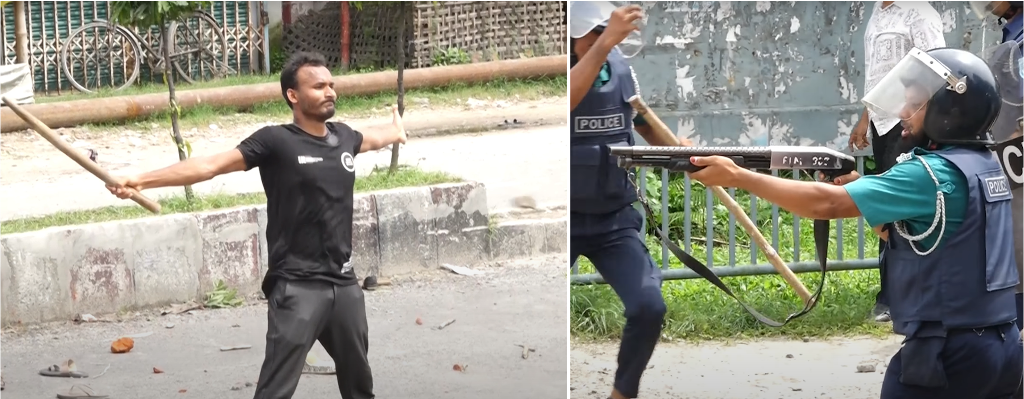}
    \caption{A collage (collected from~\cite{islam2025abu}) of widely circulated photos of a student standing with chest exposed and arms outstretched while the law enforcement agents shot him.}
    \label{fig:abusayeed}
\end{figure}

Simultaneously, the ruling party deployed both police and CL operatives against student demonstrators, inflaming tensions further. By late July, the movement had transformed into a broad pro-democracy uprising. Daily reports of deaths, injuries, and violent clashes dominated public discourse. Specific tragedies--often involving young students--were amplified across social media, galvanizing public outrage. Among those, photos, videos, and stories, such as unarmed students being killed from close range (see Figure~\ref{fig:abusayeed}) and dead bodies of the murdered students being dumped on the street by law enforcement agents, or kids playing on the roof of their own house getting shot from the sky, devastated the Bangladeshi people and their global diaspora. In response, the government launched a rolling campaign of digital suppression, shutting down platforms like Facebook, YouTube, and WhatsApp before imposing a full internet shutdown on July 18. By July 20, the protesters proposed nine specific demands, including apologies or resignations from various ministers and high-ranking officials~\cite{dailystar2024will}. The connectivity was only partially restored by July 24~\cite{zami2025bangladesh}. By then, students had effectively paralyzed transportation networks in the capital, Dhaka, prompting curfews and the deployment of additional security forces~\cite{sakib2024timeline, unohchr2024preliminary}. Between late July and early August, news reports surfaced that non-resident Bangladeshis (NRBs) were threatening to withhold remittances~\cite{hassan2024is}.

As brutality continued against the protesters, on August 3, the movement turned into a pro-democracy movement with one demand that Hasina and her cabinet members must step down~\cite{dailystar2024one}. The standoff culminated when massive demonstrations marched toward \textit{Ganabhaban}, the Prime Minister's official residence. On August 5, Hasina fled to India, bringing an end to her nearly 16-year rule~\cite{unohchr2024preliminary}. In Bangladeshi political discourse, this day is sometimes referred to as `July 36,' a symbolic extension of July that underscores how the month's unrelenting waves of protest and state violence culminated in Hasina's resignation~\cite{dailystar2025days}. To capture this historical framing, in this paper, we, like some of our participants, refer to these intertwined, fast-paced, and turbulent mobilizations as the July Movement. In the immediate aftermath, protest leaders formed an interim government led by Nobel Laureate Muhammad Yunus as Chief Advisor, a transitional arrangement that remains in place as of the time of writing this paper.

We believe that this background will help readers understand the Bangladeshi political context that led to the NRBs' remittance boycott and shaped our participants' narratives of the protests, as well as their understandings of power, resistance, and collective action.
\section{Methods}
This paper examines the 2024 remittance boycott by engaging with the lived experiences of non-resident Bangladeshis (NRBs). Having contextualized the protest within Bangladeshi sociopolitical history and existing scholarship, this section outlines the methodological approach, including the authors' positionality, recruitment, and data collection processes, and the grounded theory analysis that informed the study's findings. In line with interpretivist traditions in HCI and social computing, the methods are described with attention to how the research context, participants, and researcher positionality jointly shaped the analysis of data in this study.

\subsection{Authors' Positionality}
When researching marginalized communities, researchers' backgrounds can shape their perspectives and interpretations~\cite{schlesinger2017intersectional}. In politically sensitive contexts, it is crucial to recognize potential tensions faced by marginalized participants~\cite{liang2021embracing}. The team comprises five Bengali-speaking authors: two cisgender Bangladeshi men and one cisgender Bangladeshi woman, who have been residing in North America for approximately a decade; one cisgender Bangladeshi woman living in Bangladesh; and one cisgender Indian man, who has been living in North America for more than a decade. All but the Indian-born author maintain close family and personal connections in Bangladesh, many of whom experienced the protests firsthand. The lead and supervising authors bring extensive experience studying marginalized communities in Bangladesh, and another author served as a psychologist supporting those impacted by the July movement. While not aligned with any political ideology, all were familiar with the 2024 boycott before undertaking this project.

The first and second authors were primarily responsible for preparing the study protocols, recruiting participants, conducting interviews, translating and transcribing the data, and analyzing the results. However, they regularly discussed updates and progress of the data analyses with the rest of the research group. The authors' positionalities as academic researchers and long-term NRBs in North America have influenced the data collection and interpretation in this study.

\subsection{Recruitment}
We recruited participants through a combination of purposive and snowball sampling strategies~\cite{goodman1961snowball, suri2011purposeful} to reach a wide variety of the NRB population. We chose Facebook as the primary site to initiate our recruitment process. With 37.5\% of Bangladeshis using Facebook~\cite{statista2025leading}, the platform offers a powerful means of engaging geographically dispersed and politically diverse individuals, both within Bangladesh and across its global diaspora~\cite{sultana2022imagined}, as they create and join various Facebook groups based on different imagined communities (e.g., Bangladeshis living in Canada, the UAE, Germany, etc.). Especially during the July movement, it served as a digital hub for NRB communities and their discourse related to remittance boycotts. Particularly during the July movement, some global NRB groups were formed and became popular.

We distributed recruitment flyers for our study through both online and print channels. First, we posted the call for participants in multiple (n=14) Bangladeshi diaspora groups on Facebook, identifying them by the names of countries or major cities (e.g., Toronto, New York, Sydney) where the Bangladeshi diaspora is sizable and by the co-occurrence of keywords like ``Bangladesh," ``NRB," and ``remittance" (e.g., \textit{Remittance Fighters of Bangladesh}, \textit{Free Bangladesh–Global NRB Freedom Fighters' Network}). These groups had from tens to hundreds of thousands of members on Facebook. We contacted the groups' moderators to request approval for our posts, if they were not already approved within a few days. Their decisions determined whether and how those circulated. While these posts routinely received likes (mean = 12), comments (mean = 5), and shares, a few people expressed skepticism due to the boycott's political sensitivity. For example, some questioned how the research would benefit Bangladesh or urged us to investigate broader political corruption instead of remittance practices. While these comments were thoughtful and reflected genuine civic and political concerns, they go beyond the methodological scope of a single academic empirical study. In contrast, publicly voiced political sensitivities led many respondents to hesitate, leaving far fewer individuals willing to participate privately. Consistent with prior research~\cite{ferdous2018social}, however, we observed that when active group members or moderators tagged specific individuals, those tagged were more likely to participate.

Second, we posted 30 printed flyers in the Bangladeshi neighborhood in a major North American city. These posts and flyers outlined our recruitment criteria and included a SurveyMonkey sign-up link. For purposive sampling, we sought participants who (1) were at least 18 years of age, (2) were familiar with the remittance boycott, and (3) resided outside Bangladesh during the July movement. In the sign-up survey, we requested participants' names (optional), their preferred contact information (email or phone), and the country of their residence during the protests. We shared the interview questionnaire and consent form with our signed-up participants in advance through email or WhatsApp, based on their preferences. In those materials, we clarified that their participation was voluntary and they could participate in the interview in either Bengali or English, according to their preference, skip any questions, or end the interviews at their will. At the end of each interview, we asked if the participant knew someone who matched our recruitment criteria and might be interested in participating in our study, thereby facilitating snowball sampling. This helped us reach individuals who might not have responded to the public call.

Given the continued political uncertainty in Bangladesh, our final sample included ten NRB participants from five countries across North America, Europe, and the Middle East. Table~\ref{table:demodata} summarizes their demographics, including gender, country of residence, and educational and occupational backgrounds. We reflect on the limitations of our sampling approach and the resulting sample (e.g., limited representation of the NRBs living in the Middle East) in Section~\ref{sec:ethics_and_limitations}. Consistent with interpretivist traditions, our aim was not statistical scalability but theoretical sufficiency--having enough variation to trace the emergent phases of mobilization. We ended recruitment when additional outreach no longer yielded willing participants from the underrepresented region, despite our active recruitment efforts, and when incoming accounts started to reiterate, indicating analytical saturation regarding the NRBs' remittance boycott and its process.

\begin{table*}[!ht]
  \centering
  \caption{Participant Demographic Data}
  \label{table:demodata}
  \begin{tabular}{llllll}
    \toprule
    \textbf{Identifier} & \textbf{Age} & \textbf{Gender} & \textbf{Country of Residence} & \textbf{Education} & \textbf{Occupation Area} \\
    \midrule
    P1 & 40 & Male & United States & Masters & Data Science \\
    \hline
    P2 & 34 & Female & United States & PhD & Engineering \\
    \hline
    P3 & 26 & Male & Denmark & Masters & Business \\
    \hline
    P4 & 35 & Female & Canada & Bachelors & Data Science \\
    \hline
    P5 & 32 & Male & United States & Masters & CS/Software \\
    \hline
    P6 & 29 & Male & Germany & Bachelors & CS/Software \\
    \hline
    P7 & 41 & Male & Canada & Masters & Legal Services \\
    \hline
    P8 & 39 & Male & United States & PhD & Engineering \\
    \hline
    P9 & 34 & Male & United States & PhD & Academia \\
    \hline
    P10 & ---\footnotemark{} & Male & Saudi Arabia & --- & ---\\
    \bottomrule
  \end{tabular}
\end{table*}
\footnotetext{--- = Preferred Not to Disclose}

\subsection{Interview}
We conducted semi-structured interviews with our participants between January and March 2025. Given the geographic distance between our research team and the participants, we primarily used Microsoft Teams for the interviews, with one participant opting for a phone interview.

Our questionnaire covered topics such as participants' personal practices around remittance, perceptions of the boycott, political reflections, and experiences during the movement. Although we scheduled the interviews for 45 minutes, they often ran longer as participants elaborated on their perspectives. We recorded the interviews with the participants' consent and assured them that their responses would be de-identified and kept confidential, solely for the purpose of academic research. After transcribing all interviews, we anonymized and translated them into English for inclusion in this paper, while being mindful of sociocultural and political references.

\subsection{Data Analysis}
We employed an inductive, grounded theory-inspired approach~\cite{strauss1994grounded} for our data analysis, a method commonly used in qualitative HCI and social computing studies~\cite{das2024reimagining, houston2016values, das2021jol}. We selected it because our aim was to inductively surface the process and emergent dynamics of diasporic mobilization, which required an analytic approach oriented toward theory-building rather than descriptive patterning. We used qualitative data analysis software \textit{Atlas.ti} to code the interview transcripts. Following Strauss and Corbin's guidelines~\cite{strauss1994grounded}, we analyzed our data in three phases. In the open coding phase, we iteratively reviewed the interview transcripts and identified the abstract concepts, events, and interactions that repeatedly appeared. We generated 275 conceptually distinct open codes, such as: \textit{``remittance: important for family,"} \textit{``remittance: not important for family,"} and \textit{``participation in boycott driven by solidarity."} We identified representative quotes from the participants with the corresponding open codes. Then, we engaged in axial coding. In this phase, we grouped the related open codes into broader conceptual categories. For example, the first two aforementioned codes were combined under the axial code \textit{``remittance details and what it means for Bangladesh."} In the final selective coding phase, we condensed these axial codes into overarching themes, which we present in this paper. Because interview data are inherently contextual, our reflexive, interpretivist approach to grounded theory analysis did not require calculating inter-coder reliability scores~\cite{mcdonald2019reliability}. Moreover, it is essential to note that while our work adhered to the interpretivist tradition, the reflexive data analysis allowed patterns to emerge organically from the data, remaining sensitive to the politically charged context of the study.

\subsection{Ethical Considerations and Limitations}\label{sec:ethics_and_limitations}
We obtained approval from the University of Toronto Research Ethics Board before conducting the study. Given the political sensitivity of the topic in this study, we remained attentive to the participants' well-being during the interviews. However, despite our efforts, the study had several limitations. For example, implicit requirements such as internet access and digital communication limited accessibility to participate in our study. The sample skewed toward highly educated, professionally stable, male participants in North America and Europe, with a majority from academic or STEM backgrounds. This limits the perspectives of female participants and the significant segment of less educated, lower-income, and labor-migrant NRBs residing in the Middle East--who constitute a significant segment of the Bangladeshi diaspora. In addition, the project's politically sensitive topic limited the size of our participant pool and shaped their decision to participate and their responses. Altogether, these reflect who, across class and occupational differences, felt comfortable enough to speak up about state surveillance and political risk. Moreover, as with most qualitative studies~\cite{leung2015validity}, our findings in this study are context-specific and not intended to be generalizable; instead, they aim to examine the specific process within the defined context.
\section{Results}
In this paper, we examine the remittance boycott movement, in which non-resident Bangladeshis (NRBs) demonstrated how transnational communities can mobilize political leverage through technology. Particularly focused on their motivations and strategies--our results outline how social platforms catalyzed NRBs' shifts from passive observation to collective conviction; weak-tie networks enabled rapid coordination; financial trust systems operationalized remittance as a political tool; and adaptive strategies mitigated risks from surveillance and communication shutdowns. In this section, we describe four phases of the remittance boycott movement, beginning with its motivating factors (Phase 1) and followed by its strategic dimensions--organization, boycott execution, and barrier navigation (Phases 2-4).

\subsection{Phase 1: From Online Witnessing to Conviction}\label{sec:phase1}
Prior to the movement, NRBs held a broad spectrum of political orientations, with the majority having indifference or only limited engagement with Bangladeshi politics. Many had become desensitized or disillusioned over time, withdrawing from both online and offline political discourse due to a prevailing sense of hopelessness about the possibility of meaningful change. Some participants deliberately disengaged from political spaces as a means of self-preservation, while others restricted their involvement to minimal, passive interactions. This tendency toward detachment was evident even among those who had previously participated in political protests in the recent decade but whose interest waned following their migration abroad. They also noted that the disappointing outcomes of earlier protests significantly tempered their political optimism.

Several participants explained that the unsatisfactory outcome of past protests was a key reason for not feeling immediate hope, which had reinforced a pervasive belief--particularly among those strongly opposed to AL--that the party's rule was effectively permanent and that no viable political alternatives existed. As rhetoric about non-existent alternatives spread across online discourse during the protests, some participants like P1 found hope through certain slogans used in the July movement, such as \textit{``I am/We are the alternatives."} Some participants were inspired by such narratives and believed that common people should look for alternatives to the current government, and that an average person can also rise to be the alternative that they're seeking. These participants reported being actively engaged at the beginning of the July movement.

Although the collective morale was low at the start of the movement, NRBs of various political affiliations gradually became motivated to participate in politics, which culminated in the boycott. For the disengaged group, this mindset shift went from minimal interest to deep concern, anger, fear, and a desire for action. For those who had lost hope and were minimally engaged, their opinion shifted to a firm conviction towards the movement due to a fear for their country's future. The eventual vindication of their stance against the regime renewed their collective hope.

\subsubsection{Mobilized by Affective Visuals}
The process of NRBs' political engagement during the July movement was profoundly shaped by visuals circulating across social media platforms. Participants consistently described Facebook as the primary source of information, followed by Twitter (X) and YouTube, with mainstream media referenced less frequently and often in conjunction with these platforms. Our participants reported an inter-platform flow of information, such as accessing news clips shared by trusted journalists on Facebook or discussing developments in instant messaging groups after receiving reports from on-ground contacts.

What drove this engagement was not simply the availability of information, but the emotional and moral resonance of the visuals and slogans that stirred them into action. All participants recalled moments when they paid attention to the reports of violence, particularly state- or police-sponsored attacks. The video of Abu Sayed's\footnote{We name the everyday people killed during the July movement, drawing inspiration from \#SayHerName, a social campaign led by scholars such as Kimberl\'e Crenshaw to raise awareness of Black women who have been victims of police brutality and anti-Black violence in the United States~\cite{crenshaw2015say}.} death--which captured a student activist standing with his arms outstretched before being fatally shot (previously shown in Figure~\ref{fig:abusayeed})--emerged as a pivotal turning point in the narratives of half the participants. As Participant 2 recounted:

\begin{quote}
    On 16th July, when Abu Sayed died, that to me still seems unbelievable and nightmarish, that a person was just standing there like this with arms spread, and they killed him. To me, it's unbelievable. The day he passed, that day in the evening, I was talking to my friends like, ``No, no, I don't think that happened. Did it really happen? … Because we know, people die in cross-firing, but with that, you know there is a curtain, like killing in the dead of night. In the daylight, in front of all these people... the police committed a murder, this is an absurd, surreal thought to me even today, that it really happened.
\end{quote}

While this incident particularly devastated protesters, a few more incidents of similar violence or demonstrations of force were named by NRBs as motivators. Collectively, this includes the death of a journalist (named Tahir Zaman Priyo) and subsequent mourning, footage of helpless protesters being attacked and allegedly shot at from a helicopter, the accidental death of a child (named Riya Gope), the arrest and detainment of six coordinators of the movement, and circulating videos of brutal attacks on students. While not all participants mentioned each of these incidents by name, participants repeatedly described these visuals as crossing the limits of public tolerance, even for a country accustomed to political unrest. For some, these moments represented a personal turning point, transforming political detachment or quiet cynicism into a sense of urgency, anger, and a need to act. Even the four participants, who had been positive or hopeful about the government's tenure in Bangladesh's infrastructural development, shared that, in light of the abuses during the movement, the negatives had far outweighed the positives of the regime by that point. They held the then-government responsible and demanded that the leadership be brought to justice, if not removed from power outright. Participants who had predicted that student protesters would negotiate peace with the government started believing differently, with one participant stating:

\begin{quote}
"...after all of this went down, if these coordinators go into talks, that would be betrayal with the dead. That would be betrayal with the blood they shed, with their dead bodies. You can't have any talks after this, after the death of all of these people – after this, the current government cannot stay in power by any means. Sheikh Hasina cannot stay in power, it's not possible, and why Sheikh Hasina alone, neither her nor AL."
\end{quote}

Our participants emphasized the influence of powerful visuals and citizen journalism in online communities. The daily consumption of emotionally charged content created what participants described as ``a mental album" of key tragedies emerging online every day. The information and narratives curated through the online platforms acted as their only ``window" into the on-ground events, filtering the chaos of real-time reporting into the most emotion-gripping, shareable, engaging, and discussable narratives. This does not necessarily mean the most important information reached people online; as an example, the daily death tolls of hundreds were taken partly in stride by more than half the participants, citing such events as the norm for Bangladeshi protests. Still, Abu Sayed's emotion-evoking death on camera received stronger responses from the same group. Overall, NRBs viewed, framed, and formed their beliefs about protesters' on-the-ground experiences through the lens created by social media, simultaneously forming personal opinions, feeling heightened emotions, and debating potential calls to action through this resource. A majority reported frustration with their inability to act in this critical situation, demonstrating that while online discourse during the movement sometimes skewed perceptions, provocative visuals and narratives drew intense reactions and undeniably catalyzed collective mobilization among the NRBs.

A second strand of visuals highlighted the media appearances by the then-regime leaders. Several participants reported being aggravated by comments from current AL ministers or authorities, especially those that seemed to make light of the violence, which many NRBs interpreted as provocative or as attempts to demoralize the movement through media bites. Statements such as one minister's remark about having ``enough bullets in reserve to continue shooting for five more years" deeply angered participants. As Participant 9 reflected:

\begin{quote}
    We were all talking about things like this, how dare they buy bullets with our national money and give out these threats... So, things like this really hit people, and I honestly feel that if people associated with AL didn't speak so recklessly all the time, it's possible that people wouldn't ultimately get that angry.
\end{quote}

These moments intensified feelings of guilt and moral responsibility among NRBs, many of whom were acutely aware of their economic contributions to the country through remittances. For some, staying engaged online and amplifying protest narratives became the minimum acceptable form of action--a way to channel their frustration and solidarity when physical presence on the ground was not possible. Participant 6 described this tension clearly, noting that although they lacked the courage to return home to join street protests, leveraging their online networks felt like ``a moral duty" and ``the least [he] could do."

\subsubsection{Speaking Under Surveillance and Risk Calculus}
While social media played a significant role in politically motivating the NRBs, they also faced challenges due to fears of government interference, monitoring, and control online. Seven of our participants mentioned experiencing persecution or fear of persecution solely due to their online activity, even if they had no large online following. Their fear was not only limited to being surveilled by the government, but also extended to people affiliated with AL acting as local representatives and aggressors.

The NRBs, who were very vocal or influential on social media, were reportedly informed that they are on a list of people watched by the state, due to their years of experience rallying against AL online. And the NRBs who didn't have online influence but were vocal in their personal profile shared that they carefully deleted their online activity before traveling to Bangladesh, as a safety precaution. Threats were also felt by extended families of NRBs remaining in Bangladesh, fearing consequences as a result of their non-resident relatives' online presence. A handful of participants with close connections back home reported receiving news of the protest (via WhatsApp or phone) that never made it to mainstream media, due to state interference or the media's self-censorship. A participant\footnote{Participant ID is not specified to protect their identity further.} also shared deeply personal information during our interviews that the Bangladeshi embassy within their country of residence threatened local online activists with passport cancellation and other consequences, sometimes through anonymous phone calls. Another participant\footnotemark[3] who was particularly involved both online and offline in organizing the movement commented:

\begin{quote}
    Actually, being very honest, someone from [one of the law enforcement agencies in Bangladesh] told me that I was allegedly the first to ask for this one-point demand, which was their intelligence. [laughs] ... But I don't know, of course, if that's the truth. They have some intelligence, I guess, crawling Facebook and watching Facebook very diligently. ... And if we failed, what would happen was that everyone would be persecuted. My passport would be canceled. We received those threats. ... One of the people in the national intelligence agency told me that I was on their radar. [laughs] So definitely everyone would have to suffer, will suffer, right, if it was not going to succeed. But we didn't have any alternatives, that's the thing.
\end{quote}

Heightened fears of surveillance compelled some participants to engage with political discourse during the July movement, whereas others were immobilized by anxiety over the potential repercussions of their online activities. Despite residing abroad, several participants reported experimenting with new strategies to articulate their voices online, driven by a perceived sense of duty. One immediate adaptation involved restricting their posts to morale-boosting messages or embedding their commentary within satire and analogy to evade detection. While such self-censorship arguably constrained their capacity to fully utilize the potential of social media, it enabled them to sustain a presence within these digital spaces.

Exposure to these accounts--whether presented unfiltered or encoded through satire and analogy--reshaped how NRBs conceptualized their role in the movement. They no longer perceived themselves as passive observers but as morally implicated actors wielding leverage through their remittances. This emotional and cognitive shift laid the foundation for the organizational dynamics and boycott strategies discussed in subsequent sections.

\subsection{Phase 2: Organizing Across Borders}
Finding like-minded NRBs online provided emotional relief and a sense of solidarity. Many described the discovery of these communities as empowering, transforming isolation and guilt into a shared sense of purpose. The collective identity forged in these digital spaces enabled more coordinated and impactful actions. Becoming motivated to act and have their voices heard during this movement, NRBs organized and coordinated local protests in their respective countries.

\subsubsection{Building Network through Digital Weak Ties}
NRBs rapidly formed networks across geographic and social boundaries using social media and messaging platforms. What started as informal conversations in existing groups quickly evolved into a distributed communication network capable of coordinating actions globally.

The Bangladeshi diaspora in North America and Europe started by creating new online communication channels to connect existing large NRB communities, such as city-based Facebook groups. They established contact with one another, frequently utilizing personal networks among NRBs across different cities to arrange introductions. In doing so, they relied on messaging applications, such as WhatsApp, Telegram, Messenger, and Signal--listed in order of frequency, due to the perceived security of end-to-end encryption. They also mentioned using hashtags to recruit and reach people. Upon coming across the boycott hashtag in particular, Participant 3 commented on its growth: 

\begin{quote}
    My first reaction was just that, much like the other ideas we try, I didn't know if this would work or not. Because Sheikh Hasina's government has faced several shutdowns, protests, and movements against it over the last 15 years. Several hashtags emerged from these over time. None worked, except this one. Initially, I thought that this would be unsuccessful, just like the other times. But then I saw people discussing this hashtag in different groups, ... including on YouTube. That's when, after a day, two days, three days, or even a week later, I thought to myself, ``No, maybe something really will happen this time."
\end{quote}

Our participant 10 shared that the NRBs in Bahrain, Qatar, Saudi Arabia were already connected through various political events since 2017-2018. Participant 9 elaborated on the NRB network formed similarly within the US, describing numerous communities they personally connected with: 

\begin{quote}
At that time of crisis, we were all communicating with our contacts, asking what others were doing, what was happening in the country, what the situation was, and whether they had any information. ... At that time, the [a major US metropolitan] group added us to their WhatsApp group. Eventually, it was not just social media, but also the small local communities connecting, creating a large network [of NRBs]. They might not have done it in a very organized manner, but it eventually happened. The information transfer happened from one community to another. At least within the US, I am sure this is what happened, because I personally communicated regularly with at least 6 to 7 communities. They would tell us about the demonstrations they were doing.
\end{quote}

From these quotations above, it was evident that the NRBs were quickly forming weak ties within and beyond their communities. For some participants--especially those who had seen the sizable online opposition to the students' movement and in favor of the regime--finding this kind of community was an emotional relief. They also tried to connect with media outlets and political figures, hoping that their broad but loosely tied network could gain strength and confidence from international visibility, media attention, and conversations about why they had come together.

\subsubsection{Synergies between Online and Offline Strategies}
In addition to online organization, NRBs employed a range of offline strategies to protest. Here, we will discuss how digital coordination seamlessly translated into offline actions. Offline NRB-organized activities included demonstrations at foreign government offices, letter-writing and email campaigns to Western media and prominent political figures, contacts with Bangladeshi student associations at universities abroad to speak out against the Hasina administration, and appeals to international organizations such as Amnesty International to address the movement.

Offline protest activities often involved collaborative digital work, such as developing portfolios, and research reports, and gathering other documentation as part of globe-spanning teams. Several participants described their experiences of drafting documents for international institutions while coordinating with collaborators across the United States or in different European countries. Participant 9 shared his experience compiling a portfolio of violence from the on-the-ground movement, explaining that he essentially conducted investigative journalism through the NRB network to provide verifiable information to organizations such as the United Nations. In another example, several participants mentioned seeking guidance from non-Bangladeshi communities. For instance, NRBs in the United States reached out to the Palestinian community to learn strategies for organizing diaspora protests and to identify politicians or aid groups that might be responsive. Participant 9 underscored the significance of this cross-community information exchange:

\begin{quote}
    We even spoke with the Palestinian community, asking how they protest so their voices are heard. For example, they provided us with valuable tips, such as where to speak or protest, and which Congressman is knowledgeable about Southeast Asia and will take it seriously if contacted. All of this is achieved through community engagement among small groups.
\end{quote}

These weak-tie digital networks amplified NRB actions, reinforced collective efficacy, and--through the synergy of online coordination and offline mobilization--positioned NRBs as an organized transnational political force. The next section examines how these dynamics gave rise to the remittance boycott within this network.

\subsection{Phase 3: Collectively Turning Remittance into Leverage}
Our participants described how the idea of leveraging the government's reliance on remittance could drive collective economic action was discussed across online and offline communications among the NRBs. In this section, we will explain how the idea of boycotting remittance sending originated and how the NRBs executed it through improvisation in financial practices.

\subsubsection{Ideation of Remittance Boycott}
Regardless of their direct involvement in protests, all participants generally agreed that no single, concentrated source was responsible for promoting the boycott and perceived the idea as spreading spontaneously, gaining traction in a seemingly organic manner. While a few speculated that an organized group within the protesters might have initially planned the strategy, all of our participants reported first encountering the call for remittance boycotts online through hashtag campaigns, posts in Facebook groups, mentions by YouTube personalities, and general social media activity. Most participants mentioned that they had discussed it with peers in their local NRB communities, often to clarify whether withholding remittances would harm the country's economy more broadly, rather than specifically targeting the government. Two participants pointed to possible origins in the Middle East, with one specifically identifying a Facebook group that encouraged Middle Eastern NRBs to support the boycott:

\begin{quote}
So, in the name of the group, it had ``Remittance Boycott," or ``Bangladesh Remittance Boycott." … . However, I was actually quite surprised. Because in that group, it was entirely targeting the Middle Eastern Bangladeshi workers and laborers who reside in Middle Eastern countries, it was targeting those people and posting for them specifically. I actually found out about the remittance boycott movement there for the first time. Within that group, I also saw different types of admin and moderator posts, which encouraged the remittance boycott among group members.
\end{quote}

According to Participant 10, it is plausible that the boycott first took shape in the Middle East, although its precise origin remains uncertain. Their reflections highlight contrasts in political environments and organizing strategies between the NRBs living in the Middle East and those in Western countries. Participant 10 described how casual political conversations among friends over the past few years converged into a call for action during the July movement:

\begin{quote}
    This was all over the phone for me. ... I came back from work, did my job for the day, cooked my food, and in my downtime, went out to meet friends. That's when the talks happened. After talking for one day, two days, it hit me: we really do need to act in Bangladesh.
\end{quote}

Once the concept for a remittance boycott had reached major groups of NRBs, execution still required addressing several questions. All participants expressed the sentiment that the boycott is a tangible tool with which to protest, granting NRBs a position of power and a method of expressing their opinion in a way that would truly resonate. Participants who discussed this movement with non-Bangladeshi friends reported realizing that the economic structure and backbone of Bangladesh grant NRBs this unique position, including influence over the government if they act collectively.

The basis of political action in this sort of casual gathering was also noted for participants in the West, but the tone used by those participants was more detached than the intensity felt in the Middle East. Additionally, very few participants in the West reported years of discussion with other NRBs regarding political action; even among the politically engaged, pre-movement engagement was typically with people in Bangladesh, rather than abroad.

The differences between Middle Eastern and Western NRBs can be attributed to variations in community size and levels of political engagement. Middle Eastern NRB groups were described as both larger in number and more deeply involved in Bangladeshi politics than their Western counterparts. NRB groups in the Middle East reportedly have a precedent for protesting and boycotting during past movements, dating back at least to the events of 2018. NRB connections across countries in the Middle East are fostered through massive rallies, as well as online networks, and there is recognition that they are in a position to take on-ground action more effectively than those in the West. Participant 10 reported that many NRBs living and working in Saudi Arabia went back to Bangladesh physically during this movement, and those that stayed assisted the boycott by guarding remittance booths and asking people to withhold money:

\begin{quote}
    The places where people usually went to send money to Bangladesh--like the remittance booths--were being guarded. People skipped work and stood watch at these booths, telling others not to send remittances, especially through bKash [a popular mobile banking system in Bangladesh]. In this way, remittance sent through banking channels was stopped. There was always someone on duty saying, ``Brothers, please, don't send money to Bangladesh for a little while. This is about politics, about votes. Bangladesh cannot continue like this--the country cannot be sold to someone."
\end{quote}

However, the families of many NRBs depended on these remittances, creating a dilemma--whether to stand in solidarity with the students resisting state violence in the quota reform movement or to continue supporting their families' livelihoods. In response, NRBs adapted and modified their financial practices to balance their competing obligations as supporters of their fellow citizens' rights and as primary wage earners for their families.

\subsubsection{Financial Adaptations to Maintain Family Support}
Nearly all participants also acknowledged that the boycott is a double-edged sword: its immediate impact may not be felt by those in government, but instead on the Bangladeshi economy and the people who depend on remittances to survive. For example, while Participant 5 characterized remittance as ``the government ... getting free money without doing anything," Participant 4 emphasized how it remains ``a very high-priority thing" for her family. Participant 9 became very emotional while sharing about his family's reliance on remittance. He said:

\begin{quote}
    My family wouldn't survive if I didn't come to the U.S. and if they didn't receive the remittance, to be honest. ... Or at least the future they get to have right now, they wouldn't have then.
\end{quote}

Participant 8 expressed a similar moral opposition to the boycott saying ``this money is literally significant for our community's survival" while also considering its national impact. He said it might have been acceptable since it lasted only a short time. However, if the movement continued for a longer period, which no one knew how long it might take, it would have caused economic harm to Bangladesh before affecting those in power for an extended period. Noting that non-residents are a party that would not directly experience the consequences of a remittance boycott themselves but it would cause trouble for their country, communities, and families in various ways, this participant did not support the boycott. Participant 2 explained how the remittance boycott depriving the national reserve of foreign currencies could directly harm them:

\begin{quote}
    I would be very afraid for my parents, because they're both gone into retirement so their money is all in the bank, and reserves are depleting which is connected to the banking system, what is going to happen, what will happen to their money?
\end{quote}

To sustain families while withholding remittance through formal channels, NRBs developed informal financial workarounds. Some created trusted peer-to-peer systems to transfer money without touching state-linked reserves, while others arranged loans with fixed exchange rates or relied on travelers to hand-deliver cash.

Most participants mentioned \emph{hundi} as a method of money transfer that organically gained popularity as an alternative during that time. A few took a stance against hundi irrespective of the movement, particularly citing it as a disservice to their homeland. 

\begin{quote}
    Honestly, the whole matter of sending money through hundi seemed a bit fishy to me. When people started saying, ``Those of you from North America, don't send money through banking channels, send it through us using hundi," I began to doubt their intentions. Did this mean they were trying to profit from it somehow--maybe through exchange rates or hidden gains? I knew that if you send a large sum through one person in a hundi system, it can be manipulated, especially if the person isn't fully honest. Because of that, I didn't support hundi at the time. Between July 20th and 30th, as far as I remember, there was a massive campaign promoting it.
\end{quote}

However, limited to alternative transfer during the movement, participants mentioned witnessing, accessing, or creating alternative transfer channels themselves, in the spirit of enabling the movement rather than for profit. For example, Participant 6 developed two lists of senders and receivers within his close personal network during the movement, facilitating money transfers at the current bank rate between people who trusted him.

\begin{quote}
    During the movement, I actively posted about creating this connection and set up two pools of people: one for those who needed to send money to Bangladesh and another for those who needed to bring money here [Germany]. Since I couldn't find a secure or scalable system at the time, I relied on people I personally knew and trusted. Whenever someone had a need, they would let me know, and I would connect them. To manage this, I created two documents and took on the role of linking these groups. I posted on Facebook, asking people who wanted to send money to Bangladesh to reach out--but only if I knew them personally. The entire security of the arrangement depended on their trust in me. Although it happened through Facebook, the participants I connected with were not limited to Facebook acquaintances; they were people I already knew well.
\end{quote}

Unlike Participant 6, who was able to establish a network to facilitate transactions within borders, others did not report having access to a similar, reliable source to continue sending remittances home. While Facebook was the primary site of these discussions, where posts about the boycott often served as a space to debate opponents and drew comments suggesting alternative money transfers, people with whom they interacted were not always trustworthy, and it was challenging to find those acting in good faith. For example, two participants recalled individuals in their online networks asking the public to trust them in a hundi-like transaction. The latter participant noticed that the individual tried to profit from the situation, asking a higher price per USD than the bank rate as payment for their service. Participant 9 described the situation faced by those morally opposed to alternative money transfer but needing to send remittances urgently. They reached out to NRBs traveling to Bangladesh during that time to potentially carry a 3-month remittance sum to their families in cash. As the quotation below illustrates, Participant 1 described a different strategy in which the NRBs' families took loans from well-off acquaintances, friends, and family members in Bangladesh. These individuals would pay parties within the country, and fix their rate of repayment at the current rate for USD. The participant explained this system and its organization: 

\begin{quote}
    And they gave the loan because our logic was that the dollar would certainly appreciate in value after this economic collapse in Bangladesh, so they would make more money. ``We will send you the dollar, we are fixing the amount at the rate today, and in the future, when we are paying you in dollars, you will get more in Bangladeshi Taka." So that's there as a kind of legal arrangement we made. ... It's a lot, because I told you in our group we have 60 thousand people, in the Facebook group, a public group. There were numerous WhatsApp groups; I couldn't begin to count them, as there were so many people in various groups. So, I would say I have probably interacted directly with at least 1,000 people I never knew.
\end{quote}

As is evident, even within our participants, several semi-reliable methods of alternative money transfer emerged. These adaptations reveal how technology and trust are intertwined to make economic resistance feasible, even in the absence of scalable alternatives to formal channels of currency exchange. However, none of our participants reported a reliable, scalable alternative to bank transfers or similar remittance services. There was a general understanding among participants that all legal and morally uncontroversial transaction channels would utilize the national foreign exchange reserves to reach their intended recipients. This centralized access to remittance was at once unavoidable and a point of moral contention during this movement.   

\subsection{Phase 4: Navigating Barriers and Risks}
The already tense atmosphere within the country and in NRB communities was heightened significantly due to the nationwide internet shutdown, an event perceived by participants as deeply unacceptable in a democratic nation. Every participant mentioned the blackout causing increased anxiety and fear in their non-resident communities, with only Participant 10 reporting that people in Saudi Arabia had a suspicion that a communication shutdown was inevitable. Participant 3 recollected that he ``spoke to [his] mother after 3 days of internet blackout, so had no idea what was happening meanwhile in Bangladesh. ... Nobody found out what was happening in any corner of the country except where they lived." Several participants agreed that the internet blackout specifically alienated NRB communities, and that people who were otherwise politically disinterested, such as long-term non-residents or older demographics, were now involuntarily involved as their communications with family and friends were interrupted.

In the face of such limitations, the NRB network was one of the key resources that remained online and able to transfer some information. NRBs reported immediately brainstorming other methods of applying technology to keep communication alive, particularly to get information to those on the ground. As the first response to the blackout, Participant 2 reported telling their family and friends to use their phone number if or when they receive some bandwidth, at least to confirm that they're unharmed on the streets. Participants who were more involved with leaders in the country reported trying to finance and send phones to Bangladesh, or setting up personal virtual private networks (VPN) for at least their personal acquaintances to maintain some degree of communication:  

\begin{quote}
    We were planning to buy and send US-based cell phones to Bangladesh, so the government couldn't track them. We were involved in all these activities and worked closely together. We were supporting some groups in Bangladesh that were fighting on the street. We supported online, created [promotions], and built public opinion in favor of this ``one-demand" [of the AL-government's resignation] movement.
\end{quote}

With dwindling information exchange on the ground and limited news making it out of the country, it became important for NRBs to improvise. Communications about live emergencies in various areas, such as the need for food, water, defensive measures, and medicine, are increasingly transmitted through NRB channels. Participant 9 described an incident where students were facing a de facto siege in one of their residential dormitories (commonly referred to as halls), with police and the CL forces stationed outside. He and his community of NRBs were able to reach personal connections in the police force to try to convince them to allow food and water into the hall, leading to an anonymous case of materials being left for the students. Similar stories were reported by participants who reached out to personal connections in positions of power, particularly to learn if the Armed Forces, which they believed were partially funded by remittances and partially by international peacekeeping efforts, would support the regime. As a final example, in true bootstrapping spirit, Participant 8 researched apps such as the Briar app for decentralized communication with nearby devices, becoming a source of information for their acquaintances and some protest groups on the ground.

The ability of non-residents--many of whom reside in Global North countries--to empower their peers on the ground and show support through boycotts has implications for understanding political movements in the Global South. For a country like Bangladesh with its large NRB population and reliance on remittance, the boycott, as well as a host of other NRB contributions, had a significant impact on this revolution. 
\section{Discussion}
In the discussion, we draw out three sets of implications. First, we situate our findings within the context of critical computing, showing how our study extends postcolonial perspectives by theorizing \emph{diasporic superposition}. Second, we turn to international migrant studies, where we highlight how diasporas act not only as subjects of host-country integration but also as transnational political agents capable of shaping homeland politics. Finally, we examine how remittances intersect with the moral economies of care and resistance, and what this means for the design of financial technology (FinTech).


\subsection{Implications for Postcolonial Computing: Understanding Diasporic Agency through Superposition}
A decade of CHI and social computing research has adopted the lens of postcolonial computing to situate technologies within broader conditions of power, legitimacy, and authority~\cite{das2022decolonial, das2023studying}. This powerful analytic perspective highlights how technologies designed in the Global North and the West reproduce and impose values on communities in the Global South, urging HCI research to attend to the uneven economic relations and cultural epistemologies of the involved geographical contexts~\cite{irani2010postcolonial}. However, subsequent works have shown that adopting a simple Global North/Global South binary risks obscuring more complex positionalities. Similarly, scholars focusing on transnational mediation through technology emphasized that communities often inhabit multiple, overlapping locations of privilege and marginality~\cite{das2024reimagining, zhang2025identity}. Hence, Willems highlights the importance of theorizing nestedness--the Global North and the Global South should not be understood through their relation to each other, but through their internal hierarchies and fractures~\cite{willems2014beyond}.

\begin{figure}[!ht]
    \centering
    \includegraphics[width=\linewidth]{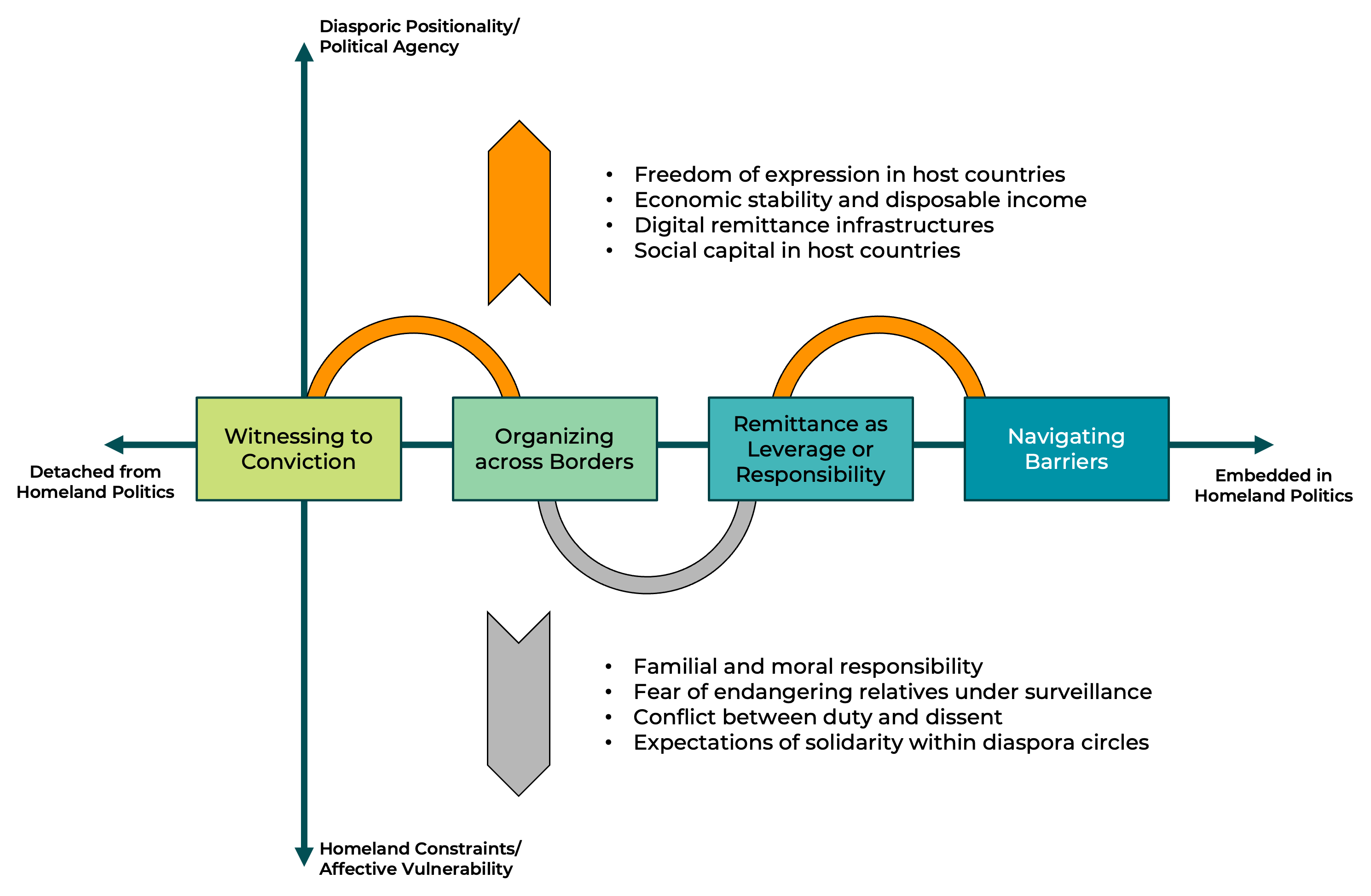}
    \caption{Diasporic superposition as oscillatory negotiation of agency and vulnerability.}
    \label{fig:superposition}
\end{figure}

The case of the 2024 remittance boycott by NRBs extends these questions. Our NRB participants occupy a complex position: originally from a Global South nation, many reside in Global North countries, working in professional or academic fields, and remain intimately tied to their homeland through family, remittances, and political attachments. While the migrants are often marginalized in their host societies, NRBs wielded significant economic and symbolic power vis-\`a-vis their homeland. They were simultaneously positioned as outsiders and insiders, privileged and vulnerable, North and South. Neither side of the binary nor the nestedness accounts for this simultaneous positionality of diasporas. Therefore, we urge critical computing scholarship to attend to their superposition in studying diasporic agency and the blurring of geographic and epistemic distinctions. We conceptualize \emph{diasporic superposition} as a dynamic condition in which migrants continuously negotiate the opposing pulls of diasporic positionality and homeland responsibility. Rather than occupying a static in-between position, diasporic actors oscillate between these poles. Through these tugging forces, their political agency in their host country is shaped by local policies, laws, and norms around freedom of expression, personal economic stability, social capital, and access to digital remittance infrastructure, whereas their vulnerability is bound by moral obligation, familial dependence and safety, and expectation of political solidarity with the majority of the diaspora. Figure~\ref{fig:superposition} illustrates this oscillation as the diasporic actors shift from detachment to active embeddedness in homeland politics through the four phases identified in our findings.

Upward arcs capture moments when diasporic positionality shapes political agency, and downward arcs represent affective and structural constraints. We dub this oscillatory dynamics as superposition--an analytical mechanism that demonstrates how diasporic actors sustain agency through informal economic network mobilization, trust-based coordination, moral discourses of care, and reflexive awareness of their dual positionality to navigate and transform structural contradictions. This formulation extends postcolonial computing (e.g., its notion of hybridity~\cite{irani2009postcolonial, bhabha2012location}) by revealing how transnational actors mobilize technologies for collective action and generate power through rhythmic negotiations of privilege and precarity across sociotechnical systems.

\subsection{Implications for Computer-Supported Cooperative Work: Expanding Affordance-Centric Models of Digital Activism}
Our study expands CSCW's understanding of digital activism by challenging affordance (i.e., analyses that focus on the actionable possibilities that platform features make available to users, such as hashtags or algorithmic amplification)-centric models that foreground visibility-making as the primary drivers of collective action~\cite{de2025good, luther2025social}. Activism and movements often emerge through contextual, place-based, and everyday practices rather than through platform features alone~\cite{crivellaro2014pool, lee2023more, das2024reimagining}. Building on this perspective, our study shows that NRB mobilization unfolded as a phased, distributed process in which affective witnessing (Phase 1) catalyzed political motivation, weak-tie formation (Phase 2) supported cross-border coordination, and improvised financial adaptations (Phase 3) transformed remittance infrastructures into political leverage. Our study further underscores that collective action is not reducible to the rapid aggregation of visible participation but emerges from the reconfiguration of long-standing sociotechnical arrangements, such as remittance channels, care infrastructures, and transnational trust networks.

This broader view of activism resonates with infrastructural analyses in CSCW, which argue that publics form through ``infrastructuring"--ongoing work that reorients existing systems and relationships toward new purposes \cite{star2006infrastructure, le2012participation}. NRBs' distributed mobilization exemplifies this: rather than relying solely on platform affordances like hashtags or trending visibility, they appropriated entrenched infrastructures of financial transfer, diaspora communication norms, and moral economies of care. The adoption of hundi or similar mechanisms through networks of trusted acquaintances for monetary transfers by some of them reflects what Asad and Le Dantec describe as ``illegitimate civic participation"--forms of activism that operate outside formal or institutionally sanctioned channels yet are deeply embedded in community infrastructures~\cite{asad2015illegitimate}. Extending CSCW insights that activism is shaped not only by on-platform actions but also by the broader, sociopolitically grounded work that sustains collective momentum~\cite{le2012participation, kow2016mediating}, our findings show how hybrid sociotechnical rhythms and rapid escalations during critical moments require CSCW research to move beyond affordance-centric analyses of visibility and coordination. Understanding diasporic mobilization demands attention to the improvised, infrastructural, and resource-leveraging practices that transform everyday systems—such as remittances—into instruments of political action.

\subsection{Implications for Social Computing: Computing for Migrants Beyond Integration}
HCI researchers studied how digital platforms mediate the everyday practices of migrant life and serve as tools for host-country integration--supporting access to services, sustaining community networks, and enabling diasporic identity work~\cite{sabie2022decade, hsiao2018technology, dosono2018identity}. While this body of work has been crucial, its analytical lens has generally been host-country-centric, focusing on how immigrants adapt to and become legible within the sociopolitical systems of their new country.

Our study offers a different perspective by foregrounding how diasporic communities influence the sociopolitical dynamics of their home countries. This paper illustrates how diasporas are not only recipients of host-country integration policies but also active agents in homeland politics. Rather than focusing on immigrants' technology adaptation, our findings show how NRBs exhibited transnational political agency by organizing and executing remittance boycott as an instrument of collective dissent against an authoritarian regime. This perspective expands HCI research on migration by emphasizing that diasporic life is not only about resettlement and integration but also about sustaining, resisting, and reshaping connections with the homeland. Prior research documents similar forms of diaspora-led mobilization in contexts such as Myanmar's political instability~\cite{williams2012changing}, Ethiopia's conflict~\cite{gregsonmigration}, and Hong Kong's pro-democracy movement~\cite{lee2023proactive}, where diasporic communities leveraged digital platforms to coordinate and amplify on-the-ground resistance across borders. While movements like those in Hong Kong and Myanmar often relied on established activist networks or exiled leadership~\cite{williams2012changing, lee2023proactive}, NRB coordination during the July movement emerged organically through weak-tie networks, without centralized organization.

Digital technologies, including remittance apps, social media, and messaging platforms, played a central role in this process. These platforms amplified calls for the boycott, circulated information about alternative transfer methods, and validated participants' choices to circumvent formal financial channels by instilling a sense of shared resistance. For social computing, these practices underscore that computing systems should not be studied solely as mediators of migrant integration, but also as a medium of transnational political action. Sociotechnical systems' verification, visibility, and surveillance directly shaped how diasporic communities could mobilize, dissent, and sustain their deliberations. By reframing diasporas as political actors rather than only subjects of integration, our study invites HCI researchers to grapple with how sociotechnical systems can both enable and constrain diasporic agency, thereby expanding migration-focused research to encompass the entanglements of technology, economy, and morality in transnational politics.

\subsection{Implications for Financial Technology and Policy: Designing for Informal Practices, Moral Decisions, and Political Agency}
Our findings suggest that FinTech cannot be separated from the moral economies shaped by obligations of kinship, ethical resistance, and visions of justice. Sending money to Bangladesh while state-sponsored violence was going on was not a neutral economic transfer for our participants: it carried moral weight as a duty of care to dependents, but also became ethically fraught when it indirectly supported the authoritarian regime's misuse of foreign reserves.

Existing FinTech systems presume remittances to be apolitical flows to be optimized for speed, cost, and compliance~\cite{narayanan2016bitcoin, maurer2012mobile}. However, the NRBs' remittance boycott showed that FinTech also mediates acts of dissent. Participants recognized that any money sent through formal banking or mobile transfer operators inevitably bolstered the very foreign reserves controlled by the regime they sought to oppose. At the same time, remittances remained indispensable for their families' survival. This paradox--the tension between supporting kin and resisting authoritarianism--surfaced as a structural challenge that current FinTech design is ill-prepared to address.

As an alternative to centralized systems, participants turned to or considered informal mechanisms such as \emph{hundi}. These informal practices, often dismissed in financial discourse as leakage or illegality, were reimagined by NRBs as means of expressing resistance and care. Similar dynamics have been observed in other contexts where sanctions, restrictions, or distrust of government shape financial practices~\cite{rohanifar2021money}. However, our study highlights the fraught nature of these alternatives: participants debated whether hundi's potential harm to the national economy outweighed its value as a tool against an oppressive regime, while also voicing concerns about profiteering or fraud by intermediaries. These deliberations show that informal practices are not simply irrational deviations from formal FinTech, but sites where migrants weigh ethics, politics, and survival.

Cryptocurrencies and blockchain-based transfers may appear to offer decentralized solutions, but our findings echo prior work showing their limited uptake among migrant communities. Trust in FinTech depends less on technical sophistication than on legibility and control over money flows~\cite{chiang2018exploring}. For NRBs, the volatility of cryptocurrencies, steep learning curve, and lack of localized technologies made them unsuitable for families dependent on stable remittance flows. As other research has found~\cite{tang2022community, rohanifar2021money}, technological solutionism that ignores sociocultural practices, political dynamics, and literacy gaps risks reproducing exclusion rather than addressing it.

In practice, NRBs improvised ad hoc trust systems, such as peer-to-peer transfers, pooled remittances, and personal courier networks, that prioritized relational accountability over institutional guarantees. Extending this to design, technologies could introduce minimal verification layers, community-reputation cues, or ephemeral coordination tools (e.g., temporary channels, low-visibility group formation, or proximity-based matching) that make such informal practices safer without over-formalizing them. These adaptations illustrate that informality is not a problem to be eliminated but a site of sociotechnical innovation which can, in some cases, also be incorporated into proprietary platforms~\cite{mwesigwa2024air}. Similarly, NRBs' practices during the boycott demonstrate how migrants appropriate existing resources to embed values of solidarity, resistance, and care.

Taken together, these findings suggest that FinTech must be reimagined as a political instrument. Future remittance platforms should not only optimize efficiency but also enable users to align financial flows with moral and political commitments. This requires monetary policies and design features that foreground transparency in how money is transferred, afford various channels of exchange rather than enforcing state-controlled pipelines, and support collective forms of action without exposing users to surveillance. Instead of assuming linear progressions from informal to formal systems, FinTech research should learn from diasporic appropriations--treating them as seeds for pragmatic, inclusive, and politically responsive design.
\section{Conclusion}
In this paper, we have explored how non-resident Bangladeshis (NRBs) mobilized by transforming remittance flows into an instrument of collective resistance during the quota reform protests in Bangladesh, which evolved into a pro-democracy uprising in July 2024. Our findings identified four phases of this mobilization--politicization through digital traces, transnational weak-tie building, the execution of a remittance boycott, and strategies for navigating surveillance and blackouts. Together, these dynamics demonstrate how diasporas act as constitutive political agents by utilizing both digital platforms and financial leverage. We extend HCI scholarship by theorizing ``diasporic superposition" in postcolonial computing and situating remittances within moral economies of care and resistance. These contributions foreground the political dimensions of the design of financial technologies and the importance of recognizing diasporas as actors in transnational collective action. Future work should focus on designing remittance transfer systems that are secure, trustworthy, transparent, resilient, and insulated from centralized surveillance and control.

\section*{Acknowledgment}
The first author is supported by a fellowship from the Institute of Health Emergencies and Pandemics at the University of Toronto. All authors thank the participants for their time and engagement in the study. We used the writing assistant software, Grammarly Premium, to improve clarity and sentence structure in our paper.

\bibliographystyle{ACM-Reference-Format}
\bibliography{sample}

@article{baxter1991bangladesh,
  title={Bangladesh in 1990: Another new beginning?},
  author={Baxter, Craig},
  journal={Asian Survey},
  volume={31},
  number={2},
  pages={146--152},
  year={1991},
  publisher={JSTOR}
}

@article{panday2010analyzing,
  title={Analyzing parliamentary election of 2008 in Bangladesh and its aftermath},
  author={Panday, PK and Jamil, I},
  journal={Towards Good Governance in South Asia, Edited by VartolaJuha, IsmoLumijarvi, Assaduzzaman Mohammed. University of Tempare, Finland},
  year={2010}
}

@article{prodip2014abolition,
  title={Abolition of non-party caretaker government system in Bangladesh: Controversy and reality},
  author={Prodip, Md Mahbub Alam and Rabbani, Golam},
  journal={Global Journal of Arts Humanities and Social Sciences},
  volume={2},
  pages={24--42},
  year={2014}
}

@article{al2013applications,
  title={Applications of e-governance towards the establishment of digital Bangladesh: Prospects and challenges},
  author={Al-Hossienie, Chowdhury Abdullah and Barua, Sato Kumer},
  journal={Journal of E-Governance},
  volume={36},
  number={3},
  pages={152--162},
  year={2013},
  publisher={SAGE Publications Sage UK: London, England}
}

@article{moniruzzaman2019regularization,
  title={Regularization of Authoritarian Democracy in Bangladesh},
  author={Moniruzzaman, M},
  journal={Elections: A Global Perspective},
  pages={57},
  year={2019},
  publisher={BoD--Books on Demand}
}

@article{riaz2021anatomy,
  title={Anatomy of a rigged election in a hybrid regime: The lessons from Bangladesh},
  author={Riaz, Ali and Parvez, Saimum},
  journal={Democratization},
  volume={28},
  number={4},
  pages={801--820},
  year={2021},
  publisher={Taylor \& Francis}
}

@article{mollah2018parliamentary,
  title={Parliamentary election and electoral violence in Bangladesh: the way forward},
  author={Mollah, Md Awal Hossain and Jahan, Rawnak},
  journal={International Journal of Law and Management},
  volume={60},
  number={2},
  pages={741--756},
  year={2018},
  publisher={Emerald Publishing Limited}
}

@article{md2025awami,
  title={Awami League's Competitive Authoritarian Rule in Bangladesh (2014-2024)},
  author={Md Saolin, Faizar},
  journal={Bangladesh (2014-2024)(May 11, 2025)},
  year={2025}
}

@techreport{surie2023glass,
  title={Glass Half Full: Civic Space and Contestation in Bangladesh, Sri Lanka and Nepal},
  author={Surie, Mandakini D and Saluja, Sumaya and Nixon, Nicola},
  year={2023},
  institution={The Asia Foundation}
}

@incollection{hossain2017significance,
  title={The significance of unruly politics in Bangladesh},
  author={Hossain, Naomi},
  booktitle={Politics and Governance in Bangladesh},
  pages={143--167},
  year={2017},
  publisher={Routledge}
}

@article{jackman2021students,
  title={Students, movements, and the threat to authoritarianism in Bangladesh},
  author={Jackman, David},
  journal={Contemporary South Asia},
  volume={29},
  number={2},
  pages={181--197},
  year={2021},
  publisher={Taylor \& Francis}
}

@incollection{tanjeem20232018,
  title={The 2018 Road Safety Protest in Bangladesh: How a Student Crowd Challenged (or Could not Challenge) the Repressive State},
  author={Tanjeem, Nafisa and Fatima, Rawshan E},
  booktitle={Young People Shaping Democratic Politics: Interrogating Inclusion, Mobilising Education},
  pages={55--81},
  year={2023},
  publisher={Springer}
}

@article{jackman2019threat,
  title={The threat of student movements in Bangladesh: Injustice, infiltrators and regime change},
  author={Jackman, David},
  year={2019},
  publisher={ESID Working Paper}
}

@article{ghosh2023conceptualizing,
  title={Conceptualizing student movements in Bangladesh post-2013: a qualitative and comparative case study of the Quota Reform Movement and the Road Safety Movement},
  author={Ghosh, Saikot Chandra},
  journal={Social Identities},
  volume={29},
  number={6},
  pages={534--554},
  year={2023},
  publisher={Taylor \& Francis}
}

@misc{sakib2024timeline,
  author = {Sakib, SM Najmus},
  title = {{TIMELINE} - {U}nprecedented student protests force {B}angladeshi {P}remier {H}asina to flee to {I}ndia},
  howpublished = {\url{https://www.aa.com.tr/en/asia-pacific/timeline-unprecedented-student-protests-force-bangladeshi-premier-hasina-to-flee-to-india/3296078}},
  year = {2024},
  note = {[Accessed 21-08-2025]},
}

@misc{unohchr2024preliminary, 
  author={United Nations Human Rights Office of the High Commissioner},
  title={Preliminary Analysis of Recent Protests and Unrest in Bangladesh},
  howpublished={\url{https://www.ohchr.org/sites/default/files/2024-08/OHCHR-Preliminary-Analysis-of-Recent-Protests-and-Unrest-in-Bangladesh-16082024_2.pdf}},
  year={2024},
  note={[Accessed 21-08-2025]}
}

@misc{das2024bangladesh,
  author = {Das, Anupreeta and Hasnat, Saif},
  title = {{B}angladesh {S}cales {B}ack {P}olicy on {P}ublic-{S}ector {H}iring {T}hat {S}parked {U}nrest},
  howpublished = {\url{https://www.nytimes.com/2024/07/21/world/asia/bangladesh-quota-court-ruling.html}},
  year = {2024},
  note = {[Accessed 21-08-2025]},
}

@misc{banka2024amidst,
  author = {Banka, Neha},
  title = {{A}midst {B}angladesh's quota reform movement, why the term ‘{R}azakar’ has provoked unprecedented fury},
  howpublished = {\url{https://indianexpress.com/article/india/bangladesh-quota-reform-movement-why-the-term-razakar-has-provoked-fury-9465108/}},
  year = {2024},
  note = {[Accessed 21-08-2025]},
}

@misc{islam2025abu,
  author = {Islam, Shadique Mahbub and Alam, Khorshed},
  title = {{H}ow {A}bu {S}ayed’s wings of freedom ignited the fire of {J}uly uprising},
  howpublished = {\url{https://www.tbsnews.net/features/panorama/how-abu-sayeeds-wings-freedom-ignited-fire-july-uprising-1189106}},
  year = {2025},
  note = {[Accessed 21-08-2025]},
}

@misc{zami2025bangladesh,
  author = {Zami, Md. Tahmid},
  title = {{B}angladesh's internet shutdown isolates citizens, disrupts business},
  howpublished = {\url{https://www.reuters.com/world/asia-pacific/bangladeshs-internet-shutdown-isolates-citizens-disrupts-business-2024-07-26/}},
  year = {2024},
  note = {[Accessed 21-08-2025]},
}

@article{liang2021embracing,
  title={Embracing Four Tensions in Human-Computer Interaction Research with Marginalized People},
  author={Liang, Calvin A and Munson, Sean A and Kientz, Julie A},
  journal={ACM Transactions on Computer-Human Interaction (TOCHI)},
  volume={28},
  number={2},
  pages={1--47},
  year={2021},
  publisher={ACM New York, NY, USA}
}

@inproceedings{schlesinger2017intersectional,
  title={Intersectional HCI: Engaging identity through gender, race, and class},
  author={Schlesinger, Ari and Edwards, W Keith and Grinter, Rebecca E},
  booktitle={Proceedings of the 2017 CHI conference on human factors in computing systems},
  pages={5412--5427},
  year={2017}
}

@misc{statista2025leading,
  author={Statista},
  title={Leading countries based on Facebook audience size},
  howpublished={\url{https://www.statista.com/statistics/268136/top-15-countries-based-on-number-of-facebook-users/}},
  year={2025},
  note={[Accessed 27-08-2025]}
}

@article{sultana2022imagined,
  title={Imagined online communities: Communionship, sovereignty, and inclusiveness in Facebook Groups},
  author={Sultana, Sharifa and Saha, Pratyasha and Hasan, Shaid and Alam, SM Raihanul and Akter, Rokeya and Islam, Md Mirajul and Arnob, Raihan Islam and Islam, AKM Najmul and Al-Ameen, Mahdi Nasrullah and Ahmed, Syed Ishtiaque},
  journal={Proceedings of the ACM on Human-Computer Interaction},
  volume={6},
  number={CSCW2},
  pages={1--29},
  year={2022},
  publisher={ACM New York, NY, USA}
}

@article{strauss1994grounded,
  title={Grounded theory methodology: An overview.},
  author={Strauss, Anselm and Corbin, Juliet},
  year={1994},
  publisher={Sage Publications, Inc}
}

@article{das2024reimagining,
  title={Reimagining Communities through Transnational Bengali Decolonial Discourse with YouTube Content Creators},
  author={Das, Dipto and Gandhi, Dhwani and Semaan, Bryan},
  journal={Proceedings of the ACM on Human-Computer Interaction},
  volume={8},
  number={CSCW2},
  pages={1--36},
  year={2024},
  publisher={ACM New York, NY, USA}
}

@inproceedings{houston2016values,
  title={Values in repair},
  author={Houston, Lara and Jackson, Steven J and Rosner, Daniela K and Ahmed, Syed Ishtiaque and Young, Meg and Kang, Laewoo},
  booktitle={Proceedings of the 2016 CHI conference on human factors in computing systems},
  pages={1403--1414},
  year={2016}
}

@article{das2021jol,
  title={" Jol" or" Pani"?: How Does Governance Shape a Platform's Identity?},
  author={Das, Dipto and {\O}sterlund, Carsten and Semaan, Bryan},
  journal={Proceedings of the ACM on Human-Computer Interaction},
  volume={5},
  number={CSCW2},
  pages={1--25},
  year={2021},
  publisher={ACM New York, NY, USA}
}

@article{mcdonald2019reliability,
  title={Reliability and inter-rater reliability in qualitative research: Norms and guidelines for CSCW and HCI practice},
  author={McDonald, Nora and Schoenebeck, Sarita and Forte, Andrea},
  journal={Proceedings of the ACM on human-computer interaction},
  volume={3},
  number={CSCW},
  pages={1--23},
  year={2019},
  publisher={ACM New York, NY, USA}
}

@article{leung2015validity,
  title={Validity, reliability, and generalizability in qualitative research},
  author={Leung, Lawrence},
  journal={Journal of family medicine and primary care},
  volume={4},
  number={3},
  pages={324--327},
  year={2015},
  publisher={Medknow}
}

@inproceedings{rohanifar2021money,
  title={Money whispers: Informality, international politics, and immigration in transnational finance},
  author={Rohanifar, Yasaman and Chandra, Priyank and Rahman, M Ataur and Ahmed, Syed Ishtiaque},
  booktitle={Proceedings of the 2021 CHI conference on human factors in computing systems},
  pages={1--14},
  year={2021}
}

@inproceedings{mwesigwa2024air,
  title={Air/time Travel: Rethinking Appropriation in Global HCI and Futures of Electronic Exchange},
  author={Mwesigwa, Daniel and Cs{\'\i}kszentmih{\'a}lyi, Christopher},
  booktitle={Proceedings of the 2024 CHI Conference on Human Factors in Computing Systems},
  pages={1--21},
  year={2024}
}

@inproceedings{chiang2018exploring,
  title={Exploring blockchain for trustful collaborations between immigrants and governments},
  author={Chiang, Chun-Wei and Betanzos, Eber and Savage, Saiph},
  booktitle={Extended Abstracts of the 2018 CHI Conference on Human Factors in Computing Systems},
  pages={1--6},
  year={2018}
}

@inproceedings{tang2022community,
  title={Community, Culture, and Capital: Exploring the Financial Practices of Older Hong Kong Immigrants},
  author={Tang, Nicole and Chandra, Priyank},
  booktitle={CHI Conference on Human Factors in Computing Systems Extended Abstracts},
  pages={1--6},
  year={2022}
}

@article{sabie2022decade,
  title={A decade of international migration research in HCI: Overview, challenges, ethics, impact, and future directions},
  author={Sabie, Dina and Ekmekcioglu, Cansu and Ahmed, Syed Ishtiaque},
  journal={ACM Transactions on Computer-Human Interaction (TOCHI)},
  volume={29},
  number={4},
  pages={1--35},
  year={2022},
  publisher={ACM New York, NY}
}

@article{srinivasan2020myths,
  title={The myths and moral economies of digital ID and mobile money in India and Myanmar},
  author={Srinivasan, Janaki and Oreglia, Elisa},
  journal={Engaging Science, Technology, and Society},
  volume={6},
  pages={215--236},
  year={2020}
}

@article{hsiao2018technology,
  title={Technology to support immigrant access to social capital and adaptation to a new country},
  author={Hsiao, Joey Chiao-Yin and Dillahunt, Tawanna R},
  journal={Proceedings of the ACM on Human-Computer Interaction},
  volume={2},
  number={CSCW},
  pages={1--21},
  year={2018},
  publisher={ACM New York, NY, USA}
}

@article{ayobi2022digital,
  title={Digital mental health and social connectedness: experiences of women from refugee backgrounds},
  author={Ayobi, Amid and Eardley, Rachel and Soubutts, Ewan and Gooberman-Hill, Rachael and Craddock, Ian and O'Kane, Aisling Ann},
  journal={Proceedings of the ACM on Human-Computer Interaction},
  volume={6},
  number={CSCW2},
  pages={1--27},
  year={2022},
  publisher={ACM New York, NY, USA}
}

@article{chen2025navigating,
  title={Navigating Immigration as an Alien: A Critical Interface Analysis of the US Citizenship and Immigration Services Website},
  author={Chen, Jianfen and Pandey, Shyam and Baniya, Sweta},
  journal={IEEE Transactions on Professional Communication},
  year={2025},
  publisher={IEEE}
}

@article{batool2024gendered,
  title={Gendered, Collectivist Journeys: Exploring Sociotechnical Adaptation Among Afghan Refugees in the USA},
  author={Batool, Amna and Dillahunt, Tawanna R and Hui, Julie and Naseem, Mustafa},
  journal={Proceedings of the ACM on Human-Computer Interaction},
  volume={8},
  number={CSCW2},
  pages={1--32},
  year={2024},
  publisher={ACM New York, NY, USA}
}

@inproceedings{dosono2018identity,
  title={Identity work as deliberation: AAPI political discourse in the 2016 US Presidential Election},
  author={Dosono, Bryan and Semaan, Bryan},
  booktitle={Proceedings of the 2018 CHI conference on human factors in computing systems},
  pages={1--12},
  year={2018}
}

@article{thompson1971moral,
  title={The moral economy of the English crowd in the eighteenth century},
  author={Thompson, Edward P},
  journal={Past \& present},
  volume={50},
  number={1},
  pages={76--136},
  year={1971},
  publisher={Oxford University Press}
}

@inproceedings{irani2010postcolonial,
  title={Postcolonial computing: a lens on design and development},
  author={Irani, Lilly and Vertesi, Janet and Dourish, Paul and Philip, Kavita and Grinter, Rebecca E},
  booktitle={Proceedings of the SIGCHI conference on human factors in computing systems},
  pages={1311--1320},
  year={2010}
}

@inproceedings{das2022decolonial,
  title={Decolonial and Postcolonial Computing Research: A Scientometric Exploration},
  author={Das, Dipto and Semaan, Bryan},
  booktitle={Companion Publication of the 2022 Conference on Computer Supported Cooperative Work and Social Computing},
  pages={168--174},
  year={2022}
}

@article{zhang2025identity,
  title={Identity Alignment and the Sociotechnical Reconfigurations of Emotional Labor in Transnational Gig-education Platforms},
  author={Zhang, Ben Zefeng and Das, Dipto and Semaan, Bryan},
  journal={Computer Supported Cooperative Work (CSCW)},
  year={2025},
  publisher={Springer}
}

@article{willems2014beyond,
  title={Beyond normative dewesternization: Examining media culture from the vantage point of the Global South},
  author={Willems, Wendy},
  journal={The Global South},
  volume={8},
  number={1},
  pages={7--23},
  year={2014},
  publisher={JSTOR}
}

@misc{horwood2021repositioning,
  title={Repositioning the importance of `migrant states' in the Global South},
  author={Horwood, Chris},
  howpublished={\url{https://mixedmigration.org/centre-stage-repositioning-the-importance-of-migrant-states-in-the-global-south/}},
  year={2021},
  note={[Accessed: 08-09-2025]}
}

@misc{paez2025top,
  title={Top Statistics on Global Migration and Migrants},
  author={Paez-Deggeller, Veronica},
  howpublished={\url{https://www.migrationpolicy.org/article/top-statistics-global-migration-migrants}},
  year={2025},
  note={[Accessed: 08-09-2025]}
}

@misc{ratha2024remittance,
  title={Remittance flows to low- and middle-income countries},
  author={Ratha, Dilip and Plaza, Sonia and Kim, Eung Ju},
  howpublished={\url{https://blogs.worldbank.org/en/peoplemove/in-2024--remittance-flows-to-low--and-middle-income-countries-ar}},
  year={2024},
  note={[Accessed: 08-09-2025]}
}

@misc{worldbank2024remittances,
  title={Remittances},
  author={World Bank},
  howpublished={\url{https://www.worldbank.org/en/topic/migration/brief/remittances-knomad}},
  year={2024},
  note={[Accessed: 08-09-2025]}
}

@article{bhattacharjee2025residual,
  title={Residual Mobilities and Religious Practices: Exploring the Experiences of the Hindu Migrants in Canada},
  author={Bhattacharjee, Ananya and Rifat, Mohammad Rashidujjaman and Das, Dipto and Haque, SM Taiabul and Ahmed, Syed Ishtiaque},
  journal={Proceedings of the ACM on Human-Computer Interaction},
  volume={9},
  number={2},
  pages={1--30},
  year={2025},
  publisher={ACM New York, NY, USA}
}

@book{brinkerhoff2009digital,
  title={Digital diasporas: Identity and transnational engagement},
  author={Brinkerhoff, Jennifer M},
  year={2009},
  publisher={Cambridge University Press}
}

@book{alonso2010diasporas,
  title={Diasporas in the new media age: Identity, politics, and community},
  author={Alonso, Andoni and Oiarzabal, Pedro},
  year={2010},
  publisher={University of Nevada Press}
}

@mastersthesis{westerhuis2022politics,
  title={Politics in Human Computer Interaction},
  author={Westerhuis, Michelle},
  year={2022},
  school={Tilburg University}
}

@article{milan2015algorithms,
  title={When algorithms shape collective action: Social media and the dynamics of cloud protesting},
  author={Milan, Stefania},
  journal={Social Media+ Society},
  volume={1},
  number={2},
  pages={2056305115622481},
  year={2015},
  publisher={SAGE Publications Sage UK: London, England}
}

@article{bilic2025digital,
  title={Digital news media as a social resilience proxy: A computational political economy perspective},
  author={Bili{\'c}, Pa{\v{s}}ko and Duki{\'c}, David and Aramba{\v{s}}i{\'c}, Lucija and Gjurkovi{\'c}, Matej and {\v{S}}najder, Jan and Furman, Ivo},
  journal={New media \& society},
  volume={27},
  number={5},
  pages={2748--2767},
  year={2025},
  publisher={SAGE Publications Sage UK: London, England}
}

@article{postmes2002collective,
  title={Collective action in the age of the Internet: Mass communication and online mobilization},
  author={Postmes, Tom and Brunsting, Suzanne},
  journal={Social science computer review},
  volume={20},
  number={3},
  pages={290--301},
  year={2002},
  publisher={Sage Publications Sage CA: Thousand Oaks, CA}
}

@article{wilson2021cross,
  title={Cross-platform Information Operations: Mobilizing Narratives \& Building Resilience through both'Big'\&'Alt'Tech},
  author={Wilson, Tom and Starbird, Kate},
  journal={Proceedings of the ACM on Human-Computer Interaction},
  volume={5},
  number={CSCW2},
  pages={1--32},
  year={2021},
  publisher={ACM New York, NY, USA}
}

@article{moss2016transnational,
  title={Transnational repression, diaspora mobilization, and the case of the Arab Spring},
  author={Moss, Dana M},
  journal={Social Problems},
  volume={63},
  number={4},
  pages={480--498},
  year={2016},
  publisher={Oxford University Press}
}

@article{rahimi2011agonistic,
  title={The agonistic social media: Cyberspace in the formation of dissent and consolidation of state power in postelection Iran},
  author={Rahimi, Babak},
  journal={The Communication Review},
  volume={14},
  number={3},
  pages={158--178},
  year={2011},
  publisher={Taylor \& Francis}
}

@article{reuter2015online,
  title={Online social media and political awareness in authoritarian regimes},
  author={Reuter, Ora John and Szakonyi, David},
  journal={British Journal of Political Science},
  volume={45},
  number={1},
  pages={29--51},
  year={2015},
  publisher={Cambridge University Press}
}

@article{ansar2024digital,
  title={Digital diaspora activism at the margins: Unfolding rohingya diaspora interactions on facebook (2017--2022)},
  author={Ansar, Anas and Maitra, Julian},
  journal={Social Media+ Society},
  volume={10},
  number={1},
  pages={20563051241228603},
  year={2024},
  publisher={SAGE Publications Sage UK: London, England}
}

@book{moss2022arab,
  title={The Arab spring abroad: Diaspora activism against authoritarian regimes},
  author={Moss, Dana M},
  year={2022},
  publisher={Cambridge University Press}
}

@article{erete2021can,
  title={I can't breathe: Reflections from Black women in CSCW and HCI},
  author={Erete, Sheena and Rankin, Yolanda A and Thomas, Jakita O},
  journal={Proceedings of the ACM on Human-Computer Interaction},
  volume={4},
  number={CSCW3},
  pages={1--23},
  year={2021},
  publisher={ACM New York, NY, USA}
}

@article{chang2023migration,
  title={Migration and financial transactions: factors influencing mobile remittance service usage in the pandemic},
  author={Chang, Wei-Lun and Benson, Vladlena},
  journal={Information Technology \& People},
  volume={36},
  number={5},
  pages={2112--2136},
  year={2023},
  publisher={Emerald Publishing Limited}
}

@article{dosono2020decolonizing,
  title={Decolonizing tactics as collective resilience: Identity work of AAPI communities on Reddit},
  author={Dosono, Bryan and Semaan, Bryan},
  journal={Proceedings of the ACM on Human-Computer interaction},
  volume={4},
  number={CSCW1},
  pages={1--20},
  year={2020},
  publisher={ACM New York, NY, USA}
}

@inproceedings{das2022understanding,
  title={Understanding the Strategies and Practices of Facebook Microcelebrities for Engaging in Sociopolitical Discourses},
  author={Das, Dipto and Islam, AKM Najmul and Haque, SM Taiabul and Vuorinen, Jukka and Ahmed, Syed Ishtiaque},
  booktitle={Proceedings of the 2022 International Conference on Information and Communication Technologies and Development},
  pages={1--19},
  year={2022}
}

@misc{mavis2024remittance,
  title={Remittance boycott putting significant pressure on reserves},
  author={Mavis, Meraj},
  howpublished={\url{https://www.dhakatribune.com/business/353074/remittance-boycott-putting-significant-pressure-on}},
  year={2024},
  note={[Accessed: 08-09-2025]}
}

@article{goodman1961snowball,
  title={Snowball sampling},
  author={Goodman, Leo A},
  journal={The annals of mathematical statistics},
  pages={148--170},
  year={1961},
  publisher={JSTOR}
}

@article{suri2011purposeful,
  title={Purposeful sampling in qualitative research synthesis},
  author={Suri, Harsh},
  journal={Qualitative research journal},
  volume={11},
  number={2},
  pages={63--75},
  year={2011},
  publisher={Emerald Group Publishing Limited}
}

@book{olson1971logic,
  title={The Logic of Collective Action: Public Goods and the Theory of Groups, with a new preface and appendix},
  author={Olson Jr, Mancur},
  volume={124},
  year={1971},
  publisher={harvard university press}
}

@article{li2018out,
  title={Out of site: Empowering a new approach to online boycotts},
  author={Li, Hanlin and Alarcon, Bodhi and Milkes Espinosa, Sara and Hecht, Brent},
  journal={Proceedings of the ACM on human-computer interaction},
  volume={2},
  number={CSCW},
  pages={1--28},
  year={2018},
  publisher={ACM New York, NY, USA}
}

@article{schradie2018digital,
  title={The digital activism gap: How class and costs shape online collective action},
  author={Schradie, Jen},
  journal={Social Problems},
  volume={65},
  number={1},
  pages={51--74},
  year={2018},
  publisher={Oxford University Press}
}

@article{enjolras2013social,
  title={Social media and mobilization to offline demonstrations: Transcending participatory divides?},
  author={Enjolras, Bernard and Steen-Johnsen, Kari and Wolleb{\ae}k, Dag},
  journal={New media \& society},
  volume={15},
  number={6},
  pages={890--908},
  year={2013},
  publisher={Sage Publications Sage UK: London, England}
}

@article{bennett2012logic,
  title={The logic of connective action: Digital media and the personalization of contentious politics},
  author={Bennett, W Lance and Segerberg, Alexandra},
  journal={Information, communication \& society},
  volume={15},
  number={5},
  pages={739--768},
  year={2012},
  publisher={Taylor \& Francis}
}

@inproceedings{cheng2014catalyst,
  title={Catalyst: triggering collective action with thresholds},
  author={Cheng, Justin and Bernstein, Michael},
  booktitle={Proceedings of the 17th ACM conference on Computer supported cooperative work \& social computing},
  pages={1211--1221},
  year={2014}
}

@article{hallam2016internet,
  title={How the internet can overcome the collective action problem: Conditional commitment designs on Pledgebank, Kickstarter, and The Point/Groupon websites},
  author={Hallam, Roger},
  journal={Information, Communication \& Society},
  volume={19},
  number={3},
  pages={362--379},
  year={2016},
  publisher={Taylor \& Francis}
}

@article{shaw2014computer,
  title={Computer supported collective action},
  author={Shaw, Aaron and Zhang, Haoqi and Monroy-Hern{\'a}ndez, Andr{\'e}s and Munson, Sean and Hill, Benjamin Mako and Gerber, Elizabeth and Kinnaird, Peter and Minder, Patrick},
  journal={Interactions},
  volume={21},
  number={2},
  pages={74--77},
  year={2014},
  publisher={ACM New York, NY, USA}
}

@article{li2025explaining,
  title={Explaining Differential Involvement in Cross-Movement Coalitions on Social Media: the\# StopHateForProfit Campaign},
  author={Li, Yevgeniya and Bernard, Jean-Gr{\'e}goire and Luczak-Roesch, Markus},
  journal={ACM Transactions on Social Computing},
  volume={8},
  number={1-2},
  pages={1--36},
  year={2025},
  publisher={ACM New York, NY}
}

@article{barron2022quantifying,
  title={Quantifying collective identity online from self-defining hashtags},
  author={Barron, Alexander TJ and Bollen, Johan},
  journal={Scientific reports},
  volume={12},
  number={1},
  pages={15044},
  year={2022},
  publisher={Nature Publishing Group UK London}
}

@article{granovetter1973strength,
  title={The strength of weak ties},
  author={Granovetter, Mark S},
  journal={American journal of sociology},
  volume={78},
  number={6},
  pages={1360--1380},
  year={1973},
  publisher={University of Chicago Press}
}

@article{aral2016future,
  title={The future of weak ties},
  author={Aral, Sinan},
  journal={American Journal of Sociology},
  volume={121},
  number={6},
  pages={1931--1939},
  year={2016},
  publisher={University of Chicago Press Chicago, IL}
}

@inproceedings{baborska2016exploring,
  title={Exploring the Relationship Between a'Facebook Group'and Face-to-Face Interactions in'Weak-Tie'Residential Communities},
  author={Baborska-Narozny, Magdalena and Stirling, Eve and Stevenson, Fionn},
  booktitle={Proceedings of the 7th 2016 International Conference on Social Media \& Society},
  pages={1--8},
  year={2016}
}

@inproceedings{wischerath2024spreading,
  title={Spreading the Word: Exploring a Network of Mobilizing Messages in a Telegram Conspiracy Group},
  author={Wischerath, Darja and Godwin, Emily and Bocheva, Desislava and Brown, Olivia and Roscoe, Jonathan Francis and Davidson, Brittany I},
  booktitle={Extended Abstracts of the CHI Conference on Human Factors in Computing Systems},
  pages={1--8},
  year={2024}
}

@article{ponzanesi2020digital,
  title={Digital diasporas: Postcoloniality, media and affect},
  author={Ponzanesi, Sandra},
  journal={Interventions},
  volume={22},
  number={8},
  pages={977--993},
  year={2020},
  publisher={Taylor \& Francis}
}

@article{candidatu2019digital,
  title={Digital diasporas: Beyond the buzzword: Toward a relational understanding of mobility and connectivity},
  author={Candidatu, Laura and Leurs, Koen and Ponzanesi, Sandra},
  journal={The handbook of diasporas, media, and culture},
  pages={31--47},
  year={2019},
  publisher={Wiley Online Library}
}

@article{jost2018social,
  title={How social media facilitates political protest: Information, motivation, and social networks},
  author={Jost, John T and Barber{\'a}, Pablo and Bonneau, Richard and Langer, Melanie and Metzger, Megan and Nagler, Jonathan and Sterling, Joanna and Tucker, Joshua A},
  journal={Political psychology},
  volume={39},
  pages={85--118},
  year={2018},
  publisher={Wiley Online Library}
}

@article{lacroix2016social,
  title={Social remittances and the changing transnational political landscape},
  author={Lacroix, Thomas and Levitt, Peggy and Vari-Lavoisier, Ilka},
  journal={Comparative migration studies},
  volume={4},
  number={1},
  pages={16},
  year={2016},
  publisher={Springer}
}

@article{krawatzek2020two,
  title={Two centuries of flows between ‘here’and ‘there’: Political remittances and their transformative potential},
  author={Krawatzek, F{\'e}lix and M{\"u}ller-Funk, Lea},
  journal={Journal of ethnic and migration studies},
  volume={46},
  number={6},
  pages={1003--1024},
  year={2020},
  publisher={Taylor \& Francis}
}

@article{escriba2018remittances,
  title={Remittances and protest in dictatorships},
  author={Escrib{\`a}-Folch, Abel and Meseguer, Covadonga and Wright, Joseph},
  journal={American Journal of Political Science},
  volume={62},
  number={4},
  pages={889--904},
  year={2018},
  publisher={Wiley Online Library}
}

@article{lopez2025migrant,
  title={Migrant money and political unrest: Remittances and support for protest in Latin America and the Caribbean},
  author={Lopez Garcia, Ana Isabel},
  journal={International Migration},
  volume={63},
  number={3},
  pages={e13351},
  year={2025},
  publisher={Wiley Online Library}
}

@article{aher2025diaspora,
  title={Do Diaspora Remittances Reduce Protest Frequency in Fragile Democracies? A Novel Non-Linear Analysis with Threshold Effects},
  author={Aher, Shreya},
  publisher={Social Science Research Network},
  year={2025}
}

@article{flanigan2017crowdfunding,
  title={Crowdfunding and diaspora philanthropy: An integration of the literature and major concepts},
  author={Flanigan, Shawn Teresa},
  journal={VOLUNTAS: International Journal of Voluntary and Nonprofit Organizations},
  volume={28},
  number={2},
  pages={492--509},
  year={2017},
  publisher={Springer}
}

@misc{international2022crowdfunding,
  title={Crowdfunding a War: The Money behind Myanmar’s Resistance},
  author={International Crisis Group},
  howpublished={\url{https://www.crisisgroup.org/sites/default/files/2022-12/328-myanmars-resistance.pdf}},
  year={2022}
}

@misc{borkena2018ethiopian,
  title={Ethiopian International Task Force calling for remittance boycott against regime in Ethiopia},
  author={Borkena},
  howpublished={\url{https://borkena.com/2018/01/01/remittance-boycott-regime-ethiopia/}},
  year={2018}
}

@article{donovan2012mobile,
  title={Mobile money for financial inclusion},
  author={Donovan, Kevin},
  journal={Information and Communications for development},
  volume={61},
  number={1},
  pages={61--73},
  year={2012}
}

@article{maurer2012mobile,
  title={Mobile money: Communication, consumption and change in the payments space},
  author={Maurer, Bill},
  journal={Journal of Development Studies},
  volume={48},
  number={5},
  pages={589--604},
  year={2012},
  publisher={Taylor \& Francis}
}

@book{narayanan2016bitcoin,
  title={Bitcoin and cryptocurrency technologies: a comprehensive introduction},
  author={Narayanan, Arvind and Bonneau, Joseph and Felten, Edward and Miller, Andrew and Goldfeder, Steven},
  year={2016},
  publisher={Princeton University Press}
}

@book{maimbo2003informal,
  title={Informal Funds Transfer Systems: An analysis of the informal hawala system},
  author={Maimbo, Mr Samuel Munzele and El Qorchi, Mr Mohammed and Wilson, Mr John F},
  year={2003},
  publisher={International Monetary Fund}
}

@inproceedings{vieweg2010microblogging,
  title={Microblogging during two natural hazards events: what twitter may contribute to situational awareness},
  author={Vieweg, Sarah and Hughes, Amanda L and Starbird, Kate and Palen, Leysia},
  booktitle={Proceedings of the SIGCHI conference on human factors in computing systems},
  pages={1079--1088},
  year={2010}
}

@inproceedings{starbird2011voluntweeters,
  title={" Voluntweeters" self-organizing by digital volunteers in times of crisis},
  author={Starbird, Kate and Palen, Leysia},
  booktitle={Proceedings of the SIGCHI conference on human factors in computing systems},
  pages={1071--1080},
  year={2011}
}

@article{li2019people,
  title={How do people change their technology use in protest?: Understanding ``Protest Users"},
  author={Li, Hanlin and Vincent, Nicholas and Tsai, Janice and Kaye, Jofish and Hecht, Brent},
  journal={Proceedings of the ACM on Human-Computer Interaction},
  volume={3},
  number={CSCW},
  pages={1--22},
  year={2019},
  publisher={ACM New York, NY, USA}
}

@inproceedings{reyes2023those,
  title={Those who are left behind: A chronicle of internet access in Cuba},
  author={Reyes Ayala, Brenda},
  booktitle={Companion Proceedings of the ACM Web Conference 2023},
  pages={610--614},
  year={2023}
}

@article{greijdanus2020psychology,
  title={The psychology of online activism and social movements: Relations between online and offline collective action},
  author={Greijdanus, Hedy and de Matos Fernandes, Carlos A and Turner-Zwinkels, Felicity and Honari, Ali and Roos, Carla A and Rosenbusch, Hannes and Postmes, Tom},
  journal={Current opinion in psychology},
  volume={35},
  pages={49--54},
  year={2020},
  publisher={Elsevier}
}

@inproceedings{al2010blogging,
  title={Blogging in a region of conflict: supporting transition to recovery},
  author={Al-Ani, Ban and Mark, Gloria and Semaan, Bryan},
  booktitle={Proceedings of the SIGCHI Conference on human factors in computing systems},
  pages={1069--1078},
  year={2010}
}

@inproceedings{das2022collaborative,
  title={Collaborative identity decolonization as reclaiming narrative agency: Identity work of Bengali communities on Quora},
  author={Das, Dipto and Semaan, Bryan},
  booktitle={Proceedings of the 2022 CHI Conference on Human Factors in Computing Systems},
  pages={1--23},
  year={2022}
}

@article{monteiro2021farmer,
  title={Farmer protests in India and the mobilization of the online diaspora on Twitter},
  author={Monteiro, Stein},
  journal={Available at SSRN 3849515},
  year={2021}
}

@article{chakrabarty2009provincializing,
  title={Provincializing Europe: postcolonial thought and historical difference-New edition},
  author={Chakrabarty, Dipesh},
  year={2009},
  publisher={Princeton university press}
}

@article{horst2008transnational,
  title={The transnational political engagements of refugees: Remittance sending practices amongst Somalis in Norway: Analysis},
  author={Horst, Cindy},
  journal={Conflict, Security \& Development},
  volume={8},
  number={3},
  pages={317--339},
  year={2008},
  publisher={Taylor \& Francis}
}

@article{chander2001diaspora,
  title={Diaspora bonds},
  author={Chander, Anupam},
  journal={NYUL Rev.},
  volume={76},
  pages={1005},
  year={2001},
  publisher={HeinOnline}
}

@misc{gevorkyan2021can,
  title={Can Diaspora Bonds Supercharge Development Investment?},
  author={Gevorkyan, Aleksandr V.},
  howpublished={\url{https://www.migrationpolicy.org/article/diaspora-bonds-supercharge-development-investment}},
  year={2021}
}

@book{saxenian2006new,
  title={The new argonauts: Regional advantage in a global economy},
  author={Saxenian, AnnaLee},
  year={2006},
  publisher={Harvard University Press}
}

@inproceedings{frohlich2022blockchain,
  title={Blockchain and cryptocurrency in human computer interaction: a systematic literature review and research agenda},
  author={Fr{\"o}hlich, Michael and Waltenberger, Franz and Trotter, Ludwig and Alt, Florian and Schmidt, Albrecht},
  booktitle={Proceedings of the 2022 ACM Designing Interactive Systems Conference},
  pages={155--177},
  year={2022}
}

@article{kaewkitipong2022human,
  title={Human--computer interaction (HCI) and trust factors for the continuance intention of mobile payment services},
  author={Kaewkitipong, Laddawan and Chen, Charlie and Han, Jiangxue and Ractham, Peter},
  journal={Sustainability},
  volume={14},
  number={21},
  pages={14546},
  year={2022},
  publisher={MDPI}
}

@article{bitrian2021making,
  title={Making finance fun: the gamification of personal financial management apps},
  author={Bitri{\'a}n, Paula and Buil, Isabel and Catal{\'a}n, Sara},
  journal={International journal of bank marketing},
  volume={39},
  number={7},
  pages={1310--1332},
  year={2021},
  publisher={Emerald Publishing Limited}
}

@inproceedings{chiang2017understanding,
  title={Understanding interface design and mobile money perceptions in Latin America},
  author={Chiang, Chun-Wei and Anderson, Caroline and Flores-Saviaga, Claudia and Arenas, Eduardo Jr and Colin, Felipe and Romero, Mario and Rivera-Loaiza, Cuauhtemoc and Chavez, Norma Elva and Savage, Saiph},
  booktitle={Proceedings of the 8th Latin American Conference on Human-Computer Interaction},
  pages={1--8},
  year={2017}
}

@article{centellegher2018mobile,
  title={Mobile money: Understanding and predicting its adoption and use in a developing economy},
  author={Centellegher, Simone and Miritello, Giovanna and Villatoro, Daniel and Parameshwar, Devyani and Lepri, Bruno and Oliver, Nuria},
  journal={Proceedings of the ACM on Interactive, Mobile, Wearable and Ubiquitous Technologies},
  volume={2},
  number={4},
  pages={1--18},
  year={2018},
  publisher={ACM New York, NY, USA}
}

@inproceedings{sowon2024role,
  title={The Role of User-Agent Interactions on Mobile Money Practices in Kenya and Tanzania},
  author={Sowon, Karen and Luhanga, Edith and Cranor, Lorrie Faith and Fanti, Giulia and Tucker, Conrad and Gueye, Assane},
  booktitle={2024 IEEE Symposium on Security and Privacy (SP)},
  pages={752--769},
  year={2024},
  organization={IEEE}
}

@inproceedings{dai2025envisioning,
  title={Envisioning Financial Technology Support for Older Adults Through Cognitive and Life Transitions},
  author={Dai, Jiamin and McGrenere, Joanna},
  booktitle={Proceedings of the 2025 CHI Conference on Human Factors in Computing Systems},
  pages={1--24},
  year={2025}
}

@article{smith2025gendered,
  title={The Gendered Algorithm: Navigating Financial Inclusion \& Equity in AI-facilitated Access to Credit},
  author={Smith, Genevieve},
  journal={arXiv preprint arXiv:2504.07312},
  year={2025}
}

@article{sohst2024leaving,
  title={Leaving No One Behind: Inclusive Fintech for Remittances},
  author={Sohst, Ravenna},
  journal={Migration Policy Institute},
  year={2024}
}

@inproceedings{sharma2025coinfused,
  title={Coinfused: Social Norms, Current Practices, and Perceived Risks among the Cryptocurrency Users},
  author={Sharma, Tanusree and Rahman, ATM Mizanur and Sandhi, Silvia and Wang, Yang and Shahriyar, Rifat and Haque, SM Taiabul},
  booktitle={Proceedings of the ACM SIGCAS/SIGCHI Conference on Computing and Sustainable Societies},
  pages={197--213},
  year={2025}
}

@misc{worldbankgroup2024personal,
  title={Personal remittances, received Percentages of GDP)},
  author={World Bank Group},
  howpublished={\url{https://data.worldbank.org/indicator/BX.TRF.PWKR.DT.GD.ZS?most_recent_value_desc=true}},
  year={2024},
  note={[Accessed: 10-09-2025]}
}

@misc{ifad2025reasons,
  title={15 reasons remittances matter},
  author={International Fund for Agricultural Development},
  howpublished={\url{https://www.ifad.org/en/w/explainers/15-reasons-remittances-matter}},
  year={2025},
  note={[Accessed: 10-09-205]}
}

@book{parrenas2005children,
  title={Children of global migration: Transnational families and gendered woes},
  author={Parre{\~n}as, Rhacel Salazar},
  year={2005},
  publisher={Stanford University Press}
}

@misc{hassan2024is,
  title={Is This the Beginning of the End of Sheikh Hasina’s Rule?},
  author={Hassan, Zia},
  howpublished={\url{https://thediplomat.com/2024/07/is-this-the-beginning-of-the-end-of-sheikh-hasinas-rule/}},
  year={2024},
  note={[Accessed: 11-09-2025]}
}

@misc{dailystar2024one,
  title={One demand now},
  author={The Daily Star},
  howpublished={\url{https://www.thedailystar.net/news/bangladesh/news/one-demand-now-3668981}},
  year={2024},
  note={[Accessed: 11-09-2025]}
}

@misc{dailystar2024will,
  title={Will consider talks if nine demands met},
  author={The Daily Star},
  howpublished={\url{https://www.thedailystar.net/news/bangladesh/news/will-consider-talks-if-nine-demands-met-3660651}},
  year={2024},
  note={[Accessed: 11-09-2025]}
}

@misc{dailystar2025days,
  title={36 Days of July},
  author={The Daily Star},
  howpublished={\url{https://july36.thedailystar.net/}},
  year={2025},
  note={[Accessed: 11-09-2025]}
}

@article{martin2009hundi,
  title={Hundi/hawala: The problem of definition1},
  author={Martin, Marina},
  journal={Modern Asian Studies},
  volume={43},
  number={4},
  pages={909--937},
  year={2009},
  publisher={Cambridge University Press}
}

@article{crenshaw2015say,
  title={Say her name: Resisting police brutality against black women},
  author={Crenshaw, Kimberl{\'e} W and Ritchie, Andrea J and Anspach, Rachel and Gilmer, Rachel and Harris, Luke},
  year={2015}
}

@inproceedings{irani2009postcolonial,
  title={Postcolonial interculturality},
  author={Irani, Lilly C and Dourish, Paul},
  booktitle={Proceedings of the 2009 international workshop on Intercultural collaboration},
  pages={249--252},
  year={2009}
}

@book{bhabha2012location,
  title={The location of culture},
  author={Bhabha, Homi K},
  year={2012},
  publisher={routledge}
}

@article{lee2023proactive,
  title={Proactive internationalization and diaspora mobilization in a networked movement: The case of Hong Kong’s Anti-Extradition Bill protests},
  author={Lee, Francis LF},
  journal={Social Movement Studies},
  volume={22},
  number={2},
  pages={232--249},
  year={2023},
  publisher={Taylor \& Francis}
}

@article{williams2012changing,
  title={Changing Burma from without: Political activism among the Burmese diaspora},
  author={Williams, David C},
  journal={Indiana Journal of Global Legal Studies},
  volume={19},
  number={1},
  pages={121--142},
  year={2012},
  publisher={JSTOR}
}

@article{gregsonmigration,
  title={Migration and conflict in the Horn of Africa},
  author={Gregson, Jessica},
  year={2020},
  howpublished={\url{https://devblog.soas.ac.uk/ref-hornresearch/files/2020/02/Literature-review-final-130120.pdf}}
}

@incollection{rahman2014social,
  title={Social organization of Hundi: informal remittance transfer to South Asia},
  author={Rahman, Md Mizanur and Yeoh, Brenda SA},
  booktitle={Migrant Remittances in South Asia: Social, Economic and Political Implications},
  pages={88--111},
  year={2014},
  publisher={Springer}
}

@inproceedings{ferdous2018social,
  title={Social media question asking (smqa) whom do we tag and why?},
  author={Ferdous, Hasan Shahid and Das, Dipto and Choudhury, Farhana Murtaza},
  booktitle={Proceedings of the 30th Australian Conference on Computer-Human Interaction},
  pages={12--22},
  year={2018}
}

@inproceedings{luther2025social,
  title={Social Media for Activists: Reimagining Safety, Content Presentation, and Workflows},
  author={Luther, Anna Ricarda and Heuer, Hendrik and Geise, Stephanie and Haunss, Sebastian and Breiter, Andreas},
  booktitle={Proceedings of the 2025 CHI Conference on Human Factors in Computing Systems},
  pages={1--18},
  year={2025}
}

@article{de2025good,
  title={Who Is a Good Digital Activist? Exploring Social Justice Activists' Adaptation to Instagram's Algorithmic Changes},
  author={De, Ankolika and Cotter, Kelley},
  journal={Proceedings of the ACM on Human-Computer Interaction},
  volume={9},
  number={7},
  pages={1--27},
  year={2025},
  publisher={ACM New York, NY, USA}
}

@inproceedings{crivellaro2014pool,
  title={A pool of dreams: facebook, politics and the emergence of a social movement},
  author={Crivellaro, Clara and Comber, Rob and Bowers, John and Wright, Peter C and Olivier, Patrick},
  booktitle={Proceedings of the SIGCHI Conference on Human Factors in Computing Systems},
  pages={3573--3582},
  year={2014}
}

@article{lee2023more,
  title={More Than a Property: Place-based Meaning Making and Mobilization on Social Media to Resist Gentrification},
  author={Lee, Seolha and Le Dantec, Christopher A},
  journal={Proceedings of the ACM on Human-Computer Interaction},
  volume={7},
  number={CSCW1},
  pages={1--20},
  year={2023},
  publisher={ACM New York, NY, USA}
}

@article{star2006infrastructure,
  title={How to infrastructure},
  author={Star, Susan Leigh and Bowker, Geoffrey C},
  journal={Handbook of new media: Social shaping and social consequences of ICTs},
  pages={230--245},
  year={2006}
}

@inproceedings{le2012participation,
  title={Participation and publics: supporting community engagement},
  author={Le Dantec, Christopher},
  booktitle={Proceedings of the SIGCHI Conference on Human Factors in Computing Systems},
  pages={1351--1360},
  year={2012}
}

@inproceedings{asad2015illegitimate,
  title={Illegitimate civic participation: supporting community activists on the ground},
  author={Asad, Mariam and Le Dantec, Christopher A},
  booktitle={Proceedings of the 18th ACM conference on Computer Supported Cooperative Work \& social computing},
  pages={1694--1703},
  year={2015}
}

@inproceedings{kow2016mediating,
  title={Mediating the undercurrents: Using social media to sustain a social movement},
  author={Kow, Yong Ming and Kou, Yubo and Semaan, Bryan and Cheng, Waikuen},
  booktitle={Proceedings of the 2016 CHI Conference on Human Factors in Computing Systems},
  pages={3883--3894},
  year={2016}
}

@article{nova2021facebook,
  title={" Facebook Promotes More Harassment" Social Media Ecosystem, Skill and Marginalized Hijra Identity in Bangladesh},
  author={Nova, Fayika Farhat and DeVito, Michael Ann and Saha, Pratyasha and Rashid, Kazi Shohanur and Roy Turzo, Shashwata and Afrin, Sadia and Guha, Shion},
  journal={Proceedings of the ACM on Human-Computer Interaction},
  volume={5},
  number={CSCW1},
  pages={1--35},
  year={2021},
  publisher={ACM New York, NY, USA}
}

@inproceedings{nova2019online,
  title={Online sexual harassment over anonymous social media in Bangladesh},
  author={Nova, Fayika Farhat and Rifat, MD Rashidujjaman and Saha, Pratyasha and Ahmed, Syed Ishtiaque and Guha, Shion},
  booktitle={Proceedings of the Tenth International Conference on Information and Communication Technologies and Development},
  pages={1--12},
  year={2019}
}

@inproceedings{sambasivan2018privacy,
  title={" Privacy is not for me, it's for those rich women": Performative Privacy Practices on Mobile Phones by Women in South Asia},
  author={Sambasivan, Nithya and Checkley, Garen and Batool, Amna and Ahmed, Nova and Nemer, David and Gayt{\'a}n-Lugo, Laura Sanely and Matthews, Tara and Consolvo, Sunny and Churchill, Elizabeth},
  booktitle={Fourteenth Symposium on Usable Privacy and Security (SOUPS 2018)},
  pages={127--142},
  year={2018}
}

@inproceedings{sambasivan2019they,
  title={``They Don't Leave Us Alone Anywhere We Go" Gender and Digital Abuse in South Asia},
  author={Sambasivan, Nithya and Batool, Amna and Ahmed, Nova and Matthews, Tara and Thomas, Kurt and Gayt{\'a}n-Lugo, Laura Sanely and Nemer, David and Bursztein, Elie and Churchill, Elizabeth and Consolvo, Sunny},
  booktitle={proceedings of the 2019 CHI Conference on Human Factors in Computing Systems},
  pages={1--14},
  year={2019}
}

@inproceedings{das2023decolonization,
  title={Decolonization through technology and decolonization of technology},
  author={Das, Dipto},
  booktitle={Companion Proceedings of the 2023 ACM International Conference on Supporting Group Work},
  pages={51--53},
  year={2023}
}

@article{rifat2024politics,
  title={The politics of fear and the experience of Bangladeshi religious minority communities using social media platforms},
  author={Rifat, Mohammad Rashidujjaman and Das, Dipto and Podder, Arpon and Jannat, Mahiratul and Soden, Robert and Semaan, Bryan and Ahmed, Syed Ishtiaque},
  journal={Proceedings of the ACM on human-computer interaction},
  volume={8},
  number={CSCW2},
  pages={1--32},
  year={2024},
  publisher={ACM New York, NY, USA}
}

@inproceedings{das2023studying,
  title={Studying multi-dimensional marginalization of identity from decolonial and postcolonial perspectives},
  author={Das, Dipto},
  booktitle={Companion Publication of the 2023 Conference on Computer Supported Cooperative Work and Social Computing},
  pages={437--440},
  year={2023}
}

@inproceedings{das2025btpd,
  title={BTPD: A Multilingual Hand-curated Dataset of B engali T ransnational P olitical D iscourse Across Online Communities},
  author={Das, Dipto and Ahmed, Syed Ishtiaque and Guha, Shion},
  booktitle={Companion Publication of the 2025 Conference on Computer-Supported Cooperative Work and Social Computing},
  pages={188--193},
  year={2025}
}

@article{das2025auditing,
  title={Auditing Political Bias in Text Generation by GPT-4 using Sociocultural and Demographic Personas: Case of Bengali Ethnolinguistic Communities},
  author={Das, Dipto and Ahmed, Syed Ishtiaque and Guha, Shion},
  booktitle={Proceedings of the First Workshop on Benchmarks, Harmonization, Annotation, and Standardization for Human-Centric AI in Indian Languages (BHASHA)},
  year={2025}
}

@inproceedings{das2024colonial,
  title={The``Colonial Impulse" of Natural Language Processing: An Audit of Bengali Sentiment Analysis Tools and Their Identity-based Biases},
  author={Das, Dipto and Guha, Shion and Brubaker, Jed R and Semaan, Bryan},
  booktitle={Proceedings of the 2024 CHI Conference on Human Factors in Computing Systems},
  pages={1--18},
  year={2024}
}

@inproceedings{das2023toward,
  title={Toward cultural bias evaluation datasets: The case of Bengali gender, religious, and national identity},
  author={Das, Dipto and Guha, Shion and Semaan, Bryan},
  booktitle={Proceedings of the First Workshop on Cross-Cultural Considerations in NLP (C3NLP)},
  pages={68--83},
  year={2023}
}

@inproceedings{sultana2024civics,
  title={A Civics-oriented Approach to Understanding Intersectionally Marginalized Users' Experience with Hate Speech Online},
  author={Sultana, Achhiya and Das, Dipto and Alam, Saadia Binte and Shidujaman, Mohammad and Ahmed, Syed Ishtiaque},
  booktitle={Proceedings of the 13th International Conference on Information \& Communication Technologies and Development},
  pages={57--68},
  year={2024}
}

@article{das2025datasets,
  title={How do datasets, developers, and models affect biases in a low-resourced language?},
  author={Das, Dipto and Guha, Shion and Semaan, Bryan},
  journal={arXiv preprint arXiv:2506.06816},
  year={2025}
}

@inproceedings{sultana2022toleration,
  title={Toleration Factors: The Expectations of Decorum, Civility, and Certainty on Rural Social Media},
  author={Sultana, Sharifa and Akter, Rokeya and Sultana, Zinnat and Ahmed, Syed Ishtiaque},
  booktitle={Proceedings of the 2022 International Conference on Information and Communication Technologies and Development},
  pages={1--14},
  year={2022}
}

@inproceedings{hasan2021oshudh,
  title={‘Oshudh Poro’: A Mobile-Phone Application to Support Low-literate Rural Bangladeshi People’s Personal Medication Management [Poster]},
  author={Hasan, Shaid and Alam, SM Raihanul and Sultana, Sharifa and Ahmed, Syed Ishtiaque},
  booktitle={Proceedings of the 4th ACM SIGCAS Conference on Computing and Sustainable Societies},
  pages={457--461},
  year={2021}
}

\end{document}